\pdfoutput=1
\documentclass[
aps, prl, reprint, twocolumn, superscriptaddress, longbibliography, floatfix ]{revtex4-2}
\usepackage[T1]{fontenc}
\usepackage[utf8]{inputenc}
\usepackage{lmodern}
\usepackage{amsmath,amssymb,mathtools, bm}
\usepackage{graphicx}
\usepackage{xcolor}
\usepackage{multirow}
\usepackage{rotating}
\usepackage{physics}
\usepackage{braket}
\usepackage{bbold}
\usepackage{comment}
\usepackage[normalem]{ulem}
\usepackage[colorlinks=true,citecolor=blue,linkcolor=blue,urlcolor=blue]{hyperref}
\usepackage[capitalise]{cleveref}
\usepackage{orcidlink}
\Crefname{figure}{Fig.}{Figs.}
\Crefname{section}{Sec.}{Secs.}

\usepackage{cmds}
\definecolor{smoothred}{HTML}{C5232F}

\makeatletter
\AtBeginDocument{%
  \def\selectlanguage#1{}%
  \def\otherlanguage#1{}%
}

\makeatother
\newcommand{\LMU}{\affiliation{Department of Physics and Arnold Sommerfeld Center for Theoretical Physics (ASC), Ludwig Maximilian University of Munich, 80333 Munich, Germany}}
\newcommand{\MPQ}{\affiliation{Max Planck Institute of Quantum Optics, 85748 Garching, Germany}}
\newcommand{\MCQST}{\affiliation{Munich Center for Quantum Science and Technology (MCQST), 80799 Munich, Germany}}
\newcommand{\ICTP}{\affiliation{The Abdus Salam International Center for Theoretical Physics (ICTP), Strada Costiera 11, 34151 Trieste, Italy}}
\newcommand{\EPFL}{\affiliation{Institute of Physics, École Polytechnique Fédérale de Lausanne (EPFL), CH-1015 Lausanne, Switzerland \\
Center for Quantum Science and Engineering, Ecole Polytechnique Fédérale de Lausanne (EPFL), CH-1015 Lausanne, Switzerland}}
\footnotetext{These authors contributed equally.}

\footnotetext{jad.halimeh@lmu.de\\emanuele.tirrito@epfl.ch}

\begin{document}
\title{Quantum Resources in Disorder-Free Localization Dynamics of Gauge Theories}
\author{Devendra Singh Bhakuni\,\orcidlink{0000-0003-3603-183X}$^{*}$}
\ICTP 
\author{Giovanni Cataldi\,\orcidlink{0000-0002-9073-8978}$^{*}$}
\author{Jad C.~Halimeh\,\orcidlink{0000-0002-0659-7990}$^{\dagger}$}\MPQ \MCQST \LMU 
\author{Emanuele Tirrito\,\orcidlink{0000-0001-7067-1203}$^{\dagger}$}
\EPFL
\date{\today}
\begin{abstract}
Quantum-state complexity diagnostics provide valuable insight into many-body dynamics, information scrambling, and quantum computation.
Here, we investigate the real-time dynamics of quantum complexity in $1\!+\!1$-dimensional Abelian U$(1)$ and non-Abelian SU$(2)$ lattice gauge theories (LGTs), focusing on the disorder-free localization (DFL) regime.
Using stabilizer Rényi entropy, participation Rényi entropy, and fermionic non-Gaussianity as measures of complexity, we observe, for both theories, two main behaviors as a function of the gauge coupling: at intermediate values, a power-law relaxation towards saturation, consistent with observations in many-body localization, and, at sufficiently large values, an ultraslow double-logarithmic growth, which we substantiate with a configuration-space bound verified by exact counting.
Our results not only provide deeper insight into the dynamics of DFL but also highlight the role of gauge invariance in constraining quantum resources and are relevant to recent quantum simulations of LGTs.
\end{abstract}
\maketitle
\paragraph*{Introduction.}
Understanding the complexity of quantum many-body states is a central challenge at the interface of quantum information and condensed-matter physics.
While entanglement has long served as a fundamental probe of quantum correlations \cite{Amico2008EntanglementManybodySystems, Horodecki2009QuantumEntanglement, Eisert2010ColloquiumAreaLaws, Plenio2014IntroductionEntanglementTheory}, it does not fully characterize the resources required for quantum computational advantage.
Recent work has therefore focused on additional signatures of non-classicality—including anti-concentration \cite{Luitz2014UniversalBehaviorBeyond, Hangleiter2018AnticoncentrationTheoremsSchemes, Mace2019MultifractalScalingsAcross, Dalzell2022RandomQuantumCircuits, Sierant2022UniversalBehaviorMultifractality, Magni2025AnticoncentrationCliffordCircuits, Lami2025AnticoncentrationStateDesign}, nonstabilizerness (magic) \cite{Bravyi2012MagicStateDistillation, Veitch2014ResourceTheoryStabilizer, Bravyi2016TradingClassicalQuantum, Bravyi2016ImprovedClassicalSimulation, Liu2022ManyBodyQuantumMagic, Bravyi2016TradingClassicalQuantum, Wang2019QuantifyingMagicQuantum, Howard2017ApplicationResourceTheory, SonyaTarabunga2025NonstabilizernessMonotoneStabilizerness}, coherence \cite{Baumgratz2014QuantifyingCoherence, Streltsov2017ColloquiumQuantumCoherence}, and fermionic non-Gaussianity \cite{Hebenstreit2019AllPureFermionic, Lumia2024MeasurementinducedTransitionsGaussianity, Lyu2024FermionicGaussianTesting, Coffman2025MeasuringNonGaussianMagic, Sierant2026FermionicMagicResources, Tarabunga2026ComputableMeasuresFermionic}.
These quantities capture different facets of quantum complexity and can be formalized within quantum resource theories \cite{Chitambar2019QuantumResourceTheories}, which provide a systematic way to relate the structure of quantum states to their classical simulability \cite{Feynman1982SimulatingPhysicsComputers}.
\begin{figure}[!t]
  \centering
  \includegraphics[width=1\columnwidth]{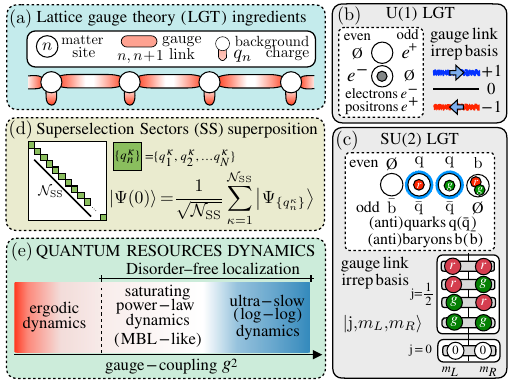}
\caption{\textbf{Disorder-free localization dynamics of quantum resources in $1\!+\!1$D LGTs.}
(a)~LGT ingredients: dynamical matter on sites $n$, gauge fields on links $(n,n{+}1)$, and static background charges $q_n$.
(b,c)~Local bases of the U$(1)$ and SU$(2)$ LGTs, with staggered fermionic matter and gauge links.
(d)~The initial states in \cref{eq_initial_state} are equal-weight superpositions of $\mathcal{N}_{\rm SS}$ superselection sectors, so the dynamics in \cref{eq_time_evolution} averages over sectors and experiences effective disorder.
(e)~Resulting resource dynamics vs.\ gauge coupling $g^2$: ergodic growth at weak coupling, MBL-like saturating power-law relaxation at intermediate $g^2$, and ultraslow double-logarithmic dynamics at the largest $g^2$. }
  \label{figpanel}
\end{figure}
In many-body quantum systems, quantum resources track the complexity of states, providing a lens on various aspects such as ground state criticality and conformal field theories\cite{White2021ConformalFieldTheories, Haug2023QuantifyingNonstabilizernessMatrix, Oliviero2022MagicstateResourceTheory, Viscardi2026InterplayEntanglementStructures, Hoshino2026StabilizerRenyiEntropy}, quantum chaos \cite{Lauchli2008SpreadingCorrelationsEntanglement, Leone2021QuantumChaosQuantum, Turkeshi2025PauliSpectrumNonstabilizerness, Turkeshi2025MagicSpreadingRandom, Jasser2025StabilizerEntropyEntanglement}, information scrambling \cite{Swingle2016MeasuringScramblingQuantum, YungerHalpern2019EntropicUncertaintyRelations, Vikram2024ProofUniversalSpeed}, thermalization \cite{Tirrito2025AnticoncentrationNonstabilizernessSpreading, Odavic2025StabilizerEntropyNonintegrable, Iannotti2026NonStabilizernessU(1)Symmetry, Tirrito2025UniversalSpreadingNonstabilizerness, Dowling2025MagicResourcesHeisenberg}, and monitored dynamics \cite{Fux2024EntanglementNonstabilizernessSeparation, Bejan2024DynamicalMagicTransitions, Tarabunga2025MagicTransitionMeasurementonly, Tirrito2025MagicPhaseTransitions, Sierant2026TheoryMagicPhase}.

Lattice gauge theories (LGTs)~\cite{Kogut1975HamiltonianFormulationWilsons, Susskind1977LatticeFermions, Wilson1974ConfinementQuarks}, that have attracted attention in quantum information science~\cite{Banuls2013TheMassSpectrum, Buyens2014MatrixProductStates, Tagliacozzo2014TensorNetworksLattice, Rico2016TensorNetworksLattice, Pichler2016RealTimeDynamics, Dalmonte2016LatticeGaugeTheorySimulations}, quantum simulation experiments~\cite{Gorg2019RealizationDensitydependentPeierls, Monroe2024ObservationStringBreaking, Gonzalez-Cuadra2025ObservationStringBreaking, Cochran2025VisualizingDynamicsCharges, Mildenberger2025ConfinementZ2Lattice, Cobos2025RealTimeDynamics2+1D, Crippa2024AnalysisConfinementString, Zhou2022ThermalizationDynamicsGauge, Wang2023InterrelatedThermalizationQuantum, Zhang2025ObservationMicroscopicConfinement, Xu2026ObservationGlueballExcitations, Joshi2026ObservationGenuine$2+1$D} and condensed matter~\cite{Wen2017ColloquiumZooQuantumtopological, Sachdev2019TopologicalOrderEmergent}, provide a particularly intriguing setting for studying complex many-body dynamics.
A defining feature of these models is gauge invariance, which constrains the dynamics of charges and eventually induces non-ergodic phenomena, such as quantum many-body scars~\cite{Turner2018weakergodicitybreaking, Surace2020LatticeGaugeTheories, Calajo2025QuantumManybodyScarring, Moudgalya2022QuantumManybodyScars, Turner2018weakergodicitybreaking, Moudgalya2018ExactExcitedStates, Iadecola2020QuantumManybodyScar, Zhao2020QuantumManyBodyScars, Aramthottil2022ScarStatesDeconfined, Jepsen2022LonglivedPhantomHelix, Bluvstein2022QuantumProcessorBased, Desaules2023ProminentQuantumManybody, Zhang2023ManybodyHilbertSpace, Budde2024QuantumManybodyScars, Serbyn2021QuantumManybodyScars, Hartse2025StabilizerScars, Desaules2023WeakErgodicityBreaking}, Hilbert-space fragmentation, and disorder-free localization (DFL) \cite{Brenes2018ManyBodyLocalizationDynamics, Osborne2023DisorderFreeLocalization$2+1$D, Ciavarella2025GenericHilbertSpace, Cataldi2026DisorderFreeLocalizationFragmentation, Gyawali2025ObservationDisorderfreeLocalization, Jeyaretnam2025HilbertSpaceFragmentation, Halimeh2022EnhancingDisorderFreeLocalization, Chakraborty2022DisorderfreeLocalizationTransition}.
While these phenomena have already been characterized in terms of spectral properties, local observables, fidelity, and entanglement entropy, a fundamental question remains: how do quantum resources spread in systems whose dynamics are governed by gauge constraints \cite{Falcao2025NonstabilizernessU1Lattice}?

In this Letter, we address this question by studying the real-time evolution of quantum resources in Abelian and non-Abelian LGTs, focusing on $(1\!+\!1)$-dimensional U$(1)$ and SU$(2)$ models.
We show that gauge invariance alone can slow down the growth of quantum complexity.
Using numerical simulations, we analyze the stabilizer Rényi entropy (SRE), the participation entropy (PE), and the fermionic non-Gaussianity (NG), uncovering an MBL-like power-law relaxation toward saturation at intermediate couplings, which then gives way to an ultraslow, double-logarithmic growth of the stabilizer and participation entropies at the largest couplings.
Our results highlight how gauge invariance affects the dynamics of quantum resources, particularly in the DFL regime, thereby bridging non-ergodic dynamics in LGTs and the resource theory of quantum complexity.

\paragraph{Quantum Resources.}
Resource theories quantify the non-classical content of quantum states by fixing a set of \emph{free states} and \emph{free operations} that cannot create the resource \cite{Chitambar2019QuantumResourceTheories}.
A monotone $\mathcal{M}$ is faithful, $\mathcal{M}(\rho)=0$ if and only if $\rho$ is free, and non-increasing under free operations $\Lambda$, $\mathcal{M}(\Lambda[\rho])\le\mathcal{M}(\rho)$.
Since each free set comes with an efficient classical simulation scheme, monotones directly gauge the hardness of simulating the dynamics.
We track three complementary monotones, associated with three distinct free sets: the SRE (nonstabilizerness), the PE (spreading over the computational basis), and the NG (correlations beyond Wick's theorem).

For a many-body system of size $L$ and Hilbert-space dimension \(D=d^{L}\) with local dimension $d$, the stabilizer Rényi entropy (SRE) of a pure state \(|\psi\rangle\) ~\cite{Leone2022StabilizerRényiEntropy, Leone2024StabilizerEntropiesAre} reads:
\begin{equation}
  {\rm SRE}_k(|\psi\rangle)
  =
  \frac{1}{1-k}
  \log\left[
    \frac{1}{D}
    \sum_{\hat P\in \mathcal P_L}
    |\langle\psi|\hat P|\psi\rangle|^{2k}
    \right],
  \label{eq:SRE}
\end{equation}
where $\mathcal{P}_L$ is the Pauli (Heisenberg--Weyl) group and purity guarantees $\sum_{\hat P}|\langle\psi|\hat P|\psi\rangle|^{2}=\sum_{\hat P}\Xi_{P}=1$.
The SRE$_k$ is thus the Rényi-$k$ entropy of the Pauli-weight distribution $\Xi_P$, offset by its stabilizer baseline $\log D$: it vanishes exactly on stabilizer states \cite{Gottesman1997StabilizerCodesQuantum, Gottesman1998TheoryFaultTolerant, Aaronson2004ImprovedSimulationStabilizer, Veitch2014ResourceTheoryStabilizer}, i.e., Clifford orbits of computational-basis product states, and for $k\ge 2$ it is a genuine magic monotone for pure states \cite{Haug2023StabilizerEntropiesNonstabilizerness, Leone2024StabilizerEntropiesAre}.
We focus on $k=2$, which is efficiently measurable via Bell/Pauli sampling~\cite{Oliviero2022MeasuringMagicQuantum, Haug2023ScalableMeasuresMagic}.

As a measure of wavefunction delocalization, we use the PE of a quantum state, defined as the Rényi-$k$ entropy of the distribution $\abs{\braket{\alpha|\psi}}^{2}$ over the computational basis $\{|\alpha\rangle\}$:
\begin{equation}
  \label{eq_pe_entropy}
  \mathrm{PE}_k(\ket{\psi})\!=\!\frac{1}{1-k} \log \qty[\sum_{\alpha=1}^{D}\abs{\braket{\alpha|\psi}}^{2k}].
\end{equation}
The PE vanishes if and only if $|\psi\rangle$ is a single basis state.
U$(1)$-symmetric Clifford unitaries act on this basis as permutations up to phases, so $\mathrm{PE}_k$ is invariant under the symmetric Clifford group $\mathcal{C}_L^{\mathrm{U(1)}}$ and quantifies magic relative to it~\cite{Liu2025QuantumAlgorithmsInverse}; dynamically, it measures how far the wave function has spread over the configuration space.
Note that $\mathrm{PE}_2$ and $\mathrm{SRE}_2$ are Rényi-2 entropies of the two distributions generated by the dynamics—$p_\alpha$ over configurations and $\Xi_{\hat P}$ over Pauli strings—a structural parallel that underlies their common late-time behavior and that we exploit in the analytical bounds below (see SM~\cite{SM}).

Finally, for fermionic systems with dynamical matter, the free states are the fermionic Gaussian states $\rho_G\!=\! c\exp\!\big(\tfrac{i}{2}\sum_{m,n=1}^{2L} \Gamma_{mn}\gamma_m\gamma_n\big)$, with $\gamma_{m}$ being Majorana operators and $\Gamma$ a real antisymmetric $2L\!\times\!2L$ matrix; free operations include the Gaussian unitaries $U_G=\exp\!\big(\tfrac12\sum_{mn}h_{mn}\gamma_m\gamma_n\big)$.
A faithful monotone is the relative entropy of non-Gaussianity~\cite{Hebenstreit2019AllPureFermionic,Lumia2024MeasurementinducedTransitionsGaussianity,Lyu2024FermionicGaussianTesting,Sierant2026FermionicMagicResources},
\begin{equation}
  \nnG(\rho)= S(\rho_G)-S(\rho),
\end{equation}
where $\rho_G$ is the Gaussian state that shares the covariance matrix of $\rho$.
NG vanishes if and only if all correlations follow from Wick's theorem.
For theories with particle number conservation, NG reduces to the binary-entropy sum over the eigenvalues $\nu_j\in[0,1]$ of the two-point function $C_{nn'}=\langle\hpsi^\dagger_n\hpsi_{n'}\rangle$.
In an LGT, the bare correlator is not gauge invariant; the physical covariance matrix dresses it with a Wilson line connecting $n$ and $n'$ (see SM~\cite{SM} for its construction in the U$(1)$ and SU$(2)$ theories).
Unlike PE and SRE, the NG is not an entropy of a dynamically generated distribution but a functional of two-point correlations only—a structural difference that anticipates its distinct late-time behavior.
Below, we use them to track how gauge invariance constrains the growth of quantum resources in the U$(1)$ and SU$(2)$ theories \cite{BasisNote}.

\begin{figure*}[t]
  \centering
  \includegraphics[width=\textwidth]{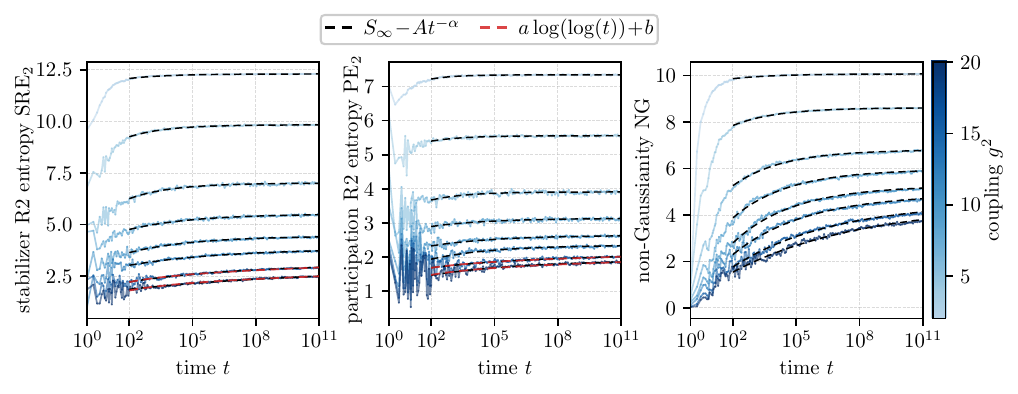}
\caption{\textbf{Time evolution of quantum resources in the U$(1)$ LGT.}
From left to right, the stabilizer Rényi-2 entropy (SRE$_2$), the participation Rényi-2 entropy (PE$_2$), and the fermionic non-Gaussianity (NG) as functions of time for vanishing mass $m\!=\!0$ (as in \cite{Brenes2018ManyBodyLocalizationDynamics}) and increasing electric coupling $g^{2}\in[2,20]$ (light to dark blue) at $L=16$.
Data and fit curves are displayed up to $t=10^{11}$, whereas the fit parameters are determined on the full window $10^2\leq t\leq10^{15}$ (see SM \cite{SM}).
Dashed black lines denote the saturating power law $X_{\infty}-At^{-\alpha}$, selected by the statistical analysis at intermediate couplings for SRE$_2$ and PE$_2$ and at every coupling for NG.
Dashed red lines denote the double-logarithmic law $a\log\log t+b$, favored by the bootstrap classification for SRE$_2$ and PE$_2$ at the two largest couplings, $g^2=16$ and $20$.}
  \label{fig_U1}
\end{figure*}

\paragraph*{Models.}
We focus on two $(1\!+\!1)$D LGTs with dynamical matter on sites $\site$ coupled to (i) Abelian U$(1)$ and (ii) non-Abelian SU$(2)$ gauge fields on links $(\site, \site\!+\!1)$.
In both cases, we employ the Kogut-Susskind formulation \cite{Kogut1975HamiltonianFormulationWilsons, Wilson1974ConfinementQuarks} with staggered fermions \cite{Susskind1977LatticeFermions} on a lattice chain of $L$ sites with unit-lattice spacing ($\lspace\!=\!1$), natural units ($\lspeed\!=\!\hbar\!=\!1$) and open boundary conditions.
For the U$(1)$ LGT, the Hamiltonian reads:
\begin{equation}
  \label{eq_U1_hamiltonian}
  \begin{split}
    \ham_{\mathrm{U(1)}}\!= & -\frac{1}{2}\sum_{\site=1}^{L-1} (\hpsi^\dagger_{\site}\Apara_{\site, \site{+}1}\hpsi_{\site{+}1}\!+\!\hc) \\
                            & \!+\!m\sum_{\site=1}^{L}(-1)^{\site}\hpsi^{\dagger}_{\site}\hpsi_{\site}
    \!+\!\frac{g^{2}}{{2}}\sum_{\site=1}^{L-1}
    \casimir_{\site, \site{+}1},
  \end{split}
\end{equation}
which describes the interaction between (spinless and flavorless) fermionic particles of mass $m$ and U$(1)$ gauge fields coupled with an electric coupling strength $g$.
In this formulation, matter fields are discretized as spinless Dirac fermions $\hpsi_{\site}$, satisfying the anti-commutation relations $\acomm*{\hpsi_{\site}}{\hpsi^{\dagger}_{\site^{\prime}}}{=} \delta_{\site,\site^{\prime}}$ and $\acomm*{\hpsi_{\site}}{\hpsi_{\site^{\prime}}}{=}0$.
As for gauge links, we adopt the basis $\qty{\ket{\ell}}_{\ell\in\mathbb{Z}}$ where the electric operator is diagonal, \idest{} $\hat{E}\ket{\ell}\!=\!\ell\ket{\ell}$, while the parallel transporter acts as a lowering operator, \idest{} $\Apara\ket{\ell}\!=\!\ket{\ell\!-\!1}$.
This definition satisfies the canonical commutation relations $[\hat{E}_{\site, \site{+}1},\Apara_{\site, \site{+}1}]\!=\!\Apara_{\site, \site{+}1}$.
Then, the U$(1)$ Gauss' law is locally manifest at each lattice site $\site$ as
\begin{equation}
  \hat{G}_{\site}\ket{\Psi}=
  \qty[\hat{E}_{\site,\site{+}1}-\hat{E}_{\site{-}1,\site}-\hat{Q}_{\site}]\ket{\Psi}
  =q_{\site}\ket{\Psi},
  \label{eq_U1_gausslaw}
\end{equation}
where $\hat{Q}_{\site}= \hpsi^{\dagger}_{\site}\hpsi_{\site}\!+\!\frac{(-1)^{\site}-1}{2}$ is the fermions charge, while $q_{\site}$ is the static background charge, corresponding to each (independent) Hamiltonian sector of $\hat{G}_{\site}$.

In one spatial dimension, exploiting U$(1)$ Gauss law in \cref{eq_U1_gausslaw} and fixing the left boundary electric field to $\hat{E}_{0,1}\!=\!\bar{E}_0$ (e.g. $\!=\!0$), we can sequentially integrate the gauge fields out (see \cite{Hamer1997SeriesExpansionsMassive, Brenes2018ManyBodyLocalizationDynamics} or \cite{SM} for more details) by expressing electric fields as a nonlocal functional of the background charges and matter fields.
With a further Jordan-Wigner transformation, one obtains an effective \emph{long-range} spin Hamiltonian whose interactions grow linearly with the lattice size:
\begin{equation}
  \begin{split}
    \ham_{\mathrm{U(1)}}^{\{q_{\site}\!\}} & \!=\!-\!w\sum_{\site=1}^{L-1}\qty(\Sp_{\site}\Sm_{\site{+}1}\!+\!\hc)
    \!+\!\frac{\mass}{2}\sum_{\site=1}^{L}(-1)^{\site}\Sz_{\site}                                                  \\
                                           & +\frac{g^{2}}{2}\sum_{\site=1}^{L-1}
    \qty[\sum_{j=1}^{\site}\qty[q_{j}+\frac{\Sz_{j}+(-1)^{j}}{2}]]^2\!+\!\rm{const.},
  \end{split}
  \label{eq_U1_hamiltonian_integrated}
\end{equation}
where $w\!=\!1/2$ is the hopping strength inherited from \cref{eq_U1_hamiltonian}.
The dynamics of \cref{eq_U1_hamiltonian_integrated} depends (through the coupling $g$) explicitly on the set of static charges $\{q_{\site}\}$ placed on the lattice and forming one of the $\mathcal{N}_{\mathrm{SS}}$ specific U$(1)$ superselection sectors (SS).
Consequently, states belonging to different superselection sectors evolve under different effective Hamiltonians $\ham^{\{q_{\site}\!\}}$.

Similarly, as for the non-Abelian case, we consider the $1\!+\!1$D SU$(2)$ Yang-Mills LGT Hamiltonian
\begin{equation}
  \label{eq_hamiltonian_SU2}
  \begin{split}
    \ham_{\mathrm{SU(2)}}= & -\frac{1}{2}\sum_{\site,\alpha, \beta}
    \qty[i\hpsi^{\dagger}_{\alpha, \site}\hat{U}^{\alpha\beta}_{\site,\site{+}1}\hpsi_{\beta,\site{+}1}+\hc]                 \\
                           & +\!\mass\!\sum_{\site,\alpha}\!(-1)^{\site} \hpsi^{\dagger}_{\alpha,\site}\hpsi_{\alpha, \site}
    \!+\!\frac{g^2}{2}\!\sum_{\site}\!\hat{E}^2_{\site,\site{+}1},
  \end{split}
\end{equation}
which describes the interaction between \emph{flavorless} SU$(2)$-color quark matter fields of mass $\mass$ and $\mathrm{SU}(2)$ gauge fields coupled with strength $g$.
In detail, quark matter fields are discretized with fermion doublets $\hpsi_{\alpha, \site}$, satisfying $\acomm*{\hpsi_{\alpha, \site}}{\hpsi^{\dagger}_{\beta,\site^{\prime}}}{=} \delta_{\site,\site^{\prime}} \delta_{\alpha, \beta}$ and $\acomm*{\hpsi_{\alpha, \site}}{\hpsi_{\beta,\site^{\prime}}}{=}0$, where $\alpha, \beta$ indices live in the fundamental $\mathrm{SU}(2)$ irreducible representation (irrep) $(\spin, \spinz)\!=\!(1/2,\pm1/2)$.
Then, the single-site matter Hilbert basis is given by:
\begin{math}
  \{{\ket{0}},\,
  \hpsi^{\dagger}_{\rla}\ket{0},\,
  \hpsi^{\dagger}_{\gla}\ket{0},\,
  \hpsi^{\dagger}_{\rla} \hpsi^{\dagger}_{\gla}\ket{0}\}
\end{math},
where $\qty{\rla,\gla}$ are shorthand notations for $\qty{\spinz\!=\!\pm1/2}$, while spin-$\spin$ is left implicit \cite{Cataldi2024Simulating2+1DSU2, Calajo2024DigitalQuantumSimulation, Calajo2025QuantumManybodyScarring, Cataldi2026DisorderFreeLocalizationFragmentation}.
For gauge link states, we adopt the chromoelectric basis $\ket{\spin, \mL, \mR}$, where $\spin\!\in\!\mathbb{N}/2$ indicates the SU$(2)$ spin irreps and $\mR,\mL\in\{-\spin,\ldots,+\spin\}$ label the states within the spin shell $\spin$.
In these bases, $\mathrm{SU}(2)$ gauge invariance is manifest at each site through generators of local rotations $\hat{G}_{\site}^{\nu}{=}(\hat{G}^{x}_{\site},\hat{G}^{y}_{\site},\hat{G}^{z}_{\site})$:
\begin{equation}
  \hat{G}^{\nu}_{\site}\ket{\Psi}\!=\!(\hat{R}_{\site{-}1,\site}^{\nu}\!+\! \hat{Q}_{\site}^{\nu}+\hat{L}_{\site,\site{+}1}^{\nu})\ket{\Psi}=q_{\site}^{\nu}\ket{\Psi},
  \label{eq_SU2_gausslaw}
\end{equation}
where
\begin{math}
  \hat{Q}^{\nu}_{\site}{=}
  \sum_{\alpha\beta}
  \hpsi^{\dagger}_{\alpha, \site}
  S^{(1/2)\nu}_{\alpha,\beta}
  \hpsi_{\beta, \site}
\end{math}
rotates the quark field at site $\site$, while $\hat{R}_{\site{-}1,\site}^{\nu}$ and $\hat{L}^{\nu}_{\site,\site{+}1}$ account for the transformation of the gauge links at its left and right \cite{Zohar2015FormulationLatticeGauge}.
Thereby, the electric energy operator
\begin{math}
  \hat{E}^2\!=\!\sum_{\nu}(\hat{R}^{\nu})^2\!=\!\sum_{\nu}(\hat{L}^{\nu})^2
\end{math}, is diagonal and coincides with the quadratic Casimir $\hat{E}^2\ket{\spin, \mL, \mR}\!=\!\spin(\spin\mathop+1)\ket{\spin, \mL, \mR}$, while the action of the parallel transporter $\NApara$ is off-diagonal and given in terms of Clebsch-Gordan coefficients \cite{Zohar2015FormulationLatticeGauge}.
Ultimately, $q_{\site}^{\nu}$ is the background charge at site $\site$ and can be expressed in an irrep basis $\ket{\spin_{q},m_{q}}$, with $\spin_{q}\!\in\!\mathbb{N}/2$.
While as in U$(1)$, gauge field integration is also accessible for SU$(2)$ \cite{Kuhn2015NonAbelianStringBreaking}, in the following, we consider the SU$(2)$ model within the minimal truncation of gauge fields and background charges, $\spin, \spin_{q}\!\in\!\qty{0, 1/2}$ (see also \cref{figpanel}(c)), which is enough to display DFL dynamics \cite{Cataldi2026DisorderFreeLocalizationFragmentation}.
Within this truncation, the SU$(2)$ Gauss law in \cref{eq_SU2_gausslaw} is solved locally via a \emph{dressed-site} approach \cite{Cataldi2024Simulating2+1DSU2, Calajo2024DigitalQuantumSimulation, Calajo2025QuantumManybodyScarring}, which yields a defermionized qudit model with a $13$-dimensional single-site basis \cite{Cataldi2026DisorderFreeLocalizationFragmentation}.

\paragraph{Initial-state preparation for DFL-dynamics.}
Following the prescription of \cite{Brenes2018ManyBodyLocalizationDynamics} for U$(1)$ and \cite{Cataldi2026DisorderFreeLocalizationFragmentation} for SU$(2)$, in order to detect DFL-dynamics, we prepare the initial state of both models in a staggered configuration of matter fields corresponding to a finite matter imbalance and gauge fields in an equal-weight superposition of all the possible values in the minimal truncation of the Casimir $\casimir$: $\norm{\casimir}\!\leq\!1$ for U$(1)$, corresponding to $\hat{E}_{\mathrm{U(1)}}\!\in\!\qty{+1, 0, -1}$, and $\norm{\casimir}\!\leq\!
3/4$ for SU$(2)$ (hardcore-gluon truncation \cite{Cataldi2024Simulating2+1DSU2, Calajo2024DigitalQuantumSimulation, Calajo2025QuantumManybodyScarring}) respectively.
Both initial states are invariant under translations by two lattice sites and can be decomposed as superpositions of U$(1)$/SU$(2)$ superselection sectors (see also \cref{figpanel}(d)):
\begin{equation}
  \label{eq_initial_state}
  \ket{\Psi(0)}
  {=}\frac{1}{\sqrt{\mathcal{N}_{\rm{SS}}}}
  \sum_{\kappa=1}^{\mathcal{N}_{\mathrm{SS}}}
  \ket{\Psi_{\{q^{\kappa}_{\site}\}}},
\end{equation}
where $\ket{\Psi_{\{q^{\kappa}_{\site}\}}}$ is a state in the $\kappa^{\mathrm{th}}$-superselection sector characterized by the specific $\kappa^{\mathrm{th}}$-configuration of background charges $\qty{q_{\site}^{\kappa}}=\qty{q_{1}^{\kappa},\dots q_{L}^{\kappa}}$, with $q_{\site}^{\kappa}\in\qty{-1, 0, 1}$ for U$(1)$, while $\spin_{{q}_{\site}^{\kappa}}\in \qty{0,1/2}$ for SU$(2)$, $\forall \kappa, \site$.
As a consequence, for an initial state $\ket{\Psi(0)}$ of the form given in \cref{eq_initial_state}, its subsequent time evolution is governed by an average over different background-charge configurations,
\begin{equation}
  \ket{\Psi(t)}\!=\!e^{-i t\ham}\ket{\Psi(0)}
  \!=\!\frac{1}{\sqrt{\mathcal{N}_{\rm{SS}}}}\!
  \sum_{\kappa=1}^{\mathcal{N}_{\rm{SS}}}e^{-it\ham^{\{q^{\kappa}_{\site}\}}}
  \!\ket{\Psi_{\{q^{\kappa}_{\site}\}}}\!,
  \label{eq_time_evolution}
\end{equation}
which effectively corresponds to a disorder average.
The effective disorder generated by averaging over superselection sectors has direct consequences on the spreading of quantum resources.

\begin{figure}[t]
  \centering
  \includegraphics[width=1\columnwidth]{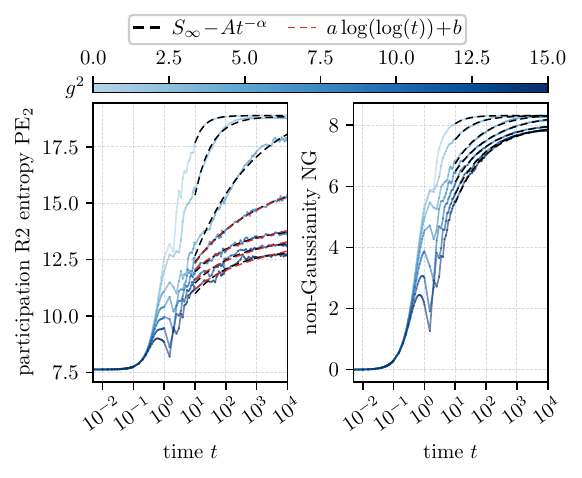}
\caption{\textbf{Time evolution of quantum resources in the SU$(2)$ LGT.}
The participation Rényi-2 entropy (PE$_2$, left) and the fermionic non-Gaussianity (NG, right) as functions of time for increasing coupling $g^{2}\!\in\![0.5,15]$ (shades of blue) at $L\!=\!12$.
After the initial transient, each trace is compared with the growth law selected by the statistical analysis (see SM): the saturating power law $X_{\infty}\!-\!At^{-\alpha}$ (dashed black) or, for PE$_2$ at strong coupling, the double-logarithmic law $a\log\log t\!+\!b$ (dashed red).
For numerical convenience, simulations are performed after rescaling the Hamiltonian in \cref{eq_hamiltonian_SU2} as $(\ham,m,g^2){\to}4\sqrt{2}(\ham,m,3 g^2/16)$ \cite{Calajo2024DigitalQuantumSimulation, Calajo2025QuantumManybodyScarring, Cataldi2026DisorderFreeLocalizationFragmentation} using the \textrm{edlgt} Python library \cite{Cataldi2026EdlgtExactDiagonalization}.}
  \label{figSU2}
\end{figure}
\begin{figure*}[t]
  \centering
  \includegraphics[width=1\textwidth]{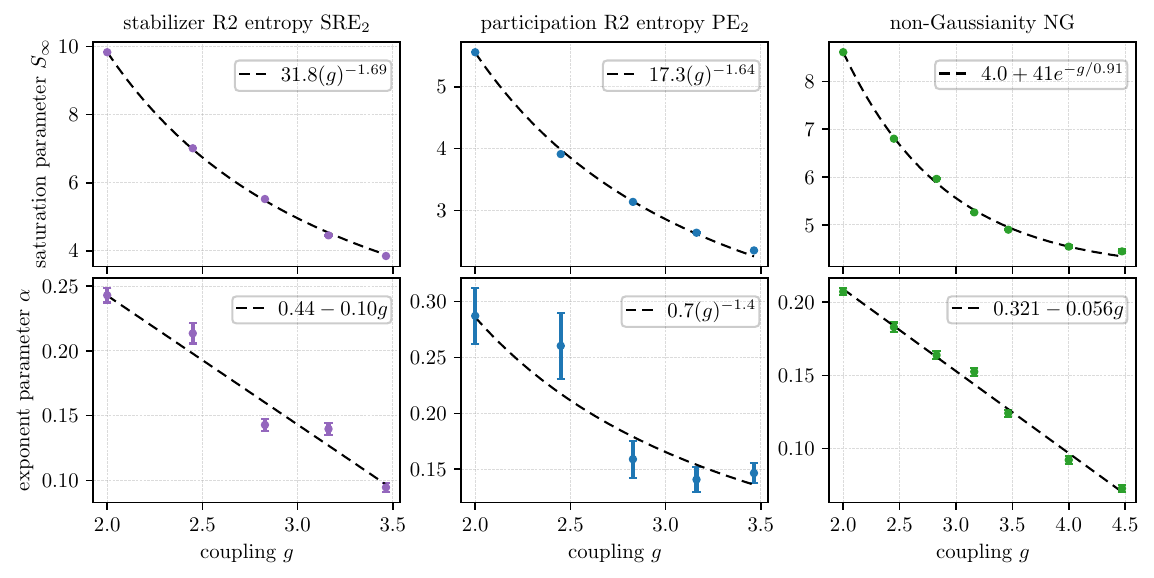}
\caption{\textbf{Scaling of the saturating-power-law parameters in the U$(1)$ LGT.}
Saturation values $X_\infty$ (top row) and relaxation exponents $\alpha$ (bottom row) of the saturating power law $X_{\infty}-At^{-\alpha}$ [\cref{eq_fit_power}] for SRE$_2$, PE$_2$, and NG (left to right), plotted as functions of the gauge coupling $g=\sqrt{g^{2}}$ and restricted to the couplings for which this law is selected by the statistical analysis (\cref{tab:time_fit_selection}).
Error bars are the covariance errors of the time-domain fits; dashed lines show the minimum-$\mathrm{AIC}$ trends among \cref{eq_parameter_trends}, reported in \cref{tab:u1_parameter_trends}.
The weakest coupling, $g^{2}=2$, lies in a distinct fast-relaxing regime and is excluded from the trend fits (see text).}
  \label{figU1_parameters}
\end{figure*}

\paragraph{Growth of quantum resources.}
We now present the time evolution of the three resource measures for different coupling strengths sketched in \cref{figpanel}(e).
\cref{fig_U1} summarizes the main dynamical results for the U$(1)$ LGT, obtained from exact simulations of \cref{eq_U1_hamiltonian_integrated} for chains of up to $L\!=\!16$ sites \cite{Cataldi2026EdlgtExactDiagonalization}, evolved according to \cref{eq_time_evolution} up to $t\!=\!10^{15}$, and with $256$ background superselection sectors entering the average.
As sketched in \cref{figpanel}, three regimes emerge upon increasing the electric coupling.
At weak coupling ($g^{2}\!\lesssim\!2$), all resources grow rapidly and equilibrate to large values, consistent with ergodic dynamics.
At intermediate couplings ($4\!\leq\!g^{2}\!\leq\!12$), the dynamics slows down and every time trace is best described by a power-law relaxation toward saturation, $X_{\mathrm{sat}}(t)\!=\!X_{\infty}\!-\!At^{-\alpha}$: the statistical model comparison detailed in the SM \cite{SM}---combining local logarithmic slopes, forward cross-validation, and a block bootstrap---selects this law with probability $P\!=\!1$ (\cref{tab:time_fit_selection}).
Such a power-law relaxation mirrors the behavior of nonstabilizerness recently reported in disorder-induced MBL \cite{Falcao2025NonstabilizernessDynamicsManyBody}, here emerging without any quenched randomness.
Correspondingly, both the saturation value $X_{\infty}$ and the relaxation exponent $\alpha$ decrease monotonically with the coupling (see \cref{figU1_parameters} in End Matter), signaling an increasingly slower and less complete relaxation.

At the largest couplings ($g^{2}\!=\!16, 20$), the late-time evolution becomes sufficiently slow that bounded and unbounded growth laws are barely distinguishable even over 13 decades in time.
For SRE$_2$ and PE$_2$, the bootstrap classification favors the unbounded double-logarithmic law $a\log\log t+b$ ($P\!\simeq\!0.90$--$0.95$), while NG remains best described by the saturating power law over the whole coupling range.
In this limit ($g^{2}\!\gg\!w$), an elementary matter hop changes the electric field on the traversed link and therefore involves an energy mismatch of order $g^2$.
One may then expect the dynamics to be governed by increasingly high-order off-resonant processes, suggesting that the perturbative depth explored up to time $t$ ---i.e., the number of hops the state has effectively performed---increases only logarithmically, $\nstar(t)\sim\log(wt)/\log(g^{2}/w)$.
The growth of PE$_2$ and SRE$_2$ then reduces to a counting problem: how many configurations can be reached within $\nstar(t)$ hops at no electric-energy cost.
A charged excitation has an $\mathcal{O}(g^{2})$ energy cost, so confinement caps the number of excitations that can move: a configuration is reached by displacing a few mobile charges, and the number $\Vres$ of accessible configurations grows \emph{polynomially} with the depth, $\Vres(\nstar)\!\sim\!\nstar^{k}$, with the exponent $k$ counting the mobile charges.
This polynomial growth is verified by exactly counting the accessible configurations with a transfer-matrix method up to $L\!=\!96$, which yields an exponent growing linearly with system size, $k\!\simeq\!0.15L$, \idest{} a finite \emph{density} of mobile charges~\cite{SM}.
Since PE$_2$ is bounded by the logarithm of this volume, and $\SRE\le2\,\PE$~\cite{SM}, both resources are capped by the ultraslow ceiling ${\sim}\,k\log\log t$, with a downward shift $-k\log\log(g^{2}/w)$ upon increasing the coupling, consistent with \cref{fig_U1}.

The SU$(2)$ dynamics in \cref{figSU2} displays a closely related coupling-dependent slowdown \cite{SreNote}.
At weak coupling, PE$_2$ grows rapidly after the initial transient and approaches a finite plateau, whereas increasing $g^{2}$ delays wave-function spreading and substantially reduces the configuration-space volume explored within the accessible time window.
Correspondingly, the statistical analysis selects a saturating power law for $g^{2}\!\leq\!2.92$ ($P_{\mathrm{sat}}=1$), while for $g^{2}\!\geq\!5.33$ it favors continued double-logarithmic growth.
The preference is strongest at $g^{2}=7.75$ ($P=0.95$) and becomes less decisive at the largest couplings, where the evolution is extremely slow.
In contrast, NG approaches a finite plateau at every coupling ($P_{\mathrm{sat}}=1$): increasing $g^{2}$ mainly delays its buildup rather than changing its asymptotic functional form.
The SU$(2)$ results therefore reproduce the separation observed in the U$(1)$ theory between the ultraslow spreading of the many-body wave function, captured by PE$_2$, and the saturation of the two-point-correlation-based NG.
Since the available SU$(2)$ time window spans only approximately three decades, however, the strong-coupling classification should be regarded as indicative rather than conclusive (see SM \cite{SM}).
In the End Matter, we complement these dynamical results with the static, equilibrium behavior of the same resources, whose finite-size analysis distinguishes the ergodic and localized regimes (\cref{fig_U1_statics}).

\paragraph{Conclusions.}
In this work, we investigated how gauge invariance constrains the dynamics of quantum resources in $(1\!+\!1)$D Abelian U$(1)$ and non-Abelian SU$(2)$ LGTs, using the stabilizer Rényi entropy, the participation entropy, and the fermionic non-Gaussianity as quantitative measures.
Quenching from translationally invariant superpositions of superselection sectors---the protocol known to induce DFL---and combining exact simulations with a systematic statistical model comparison, we identified two distinct non-ergodic regimes.
At intermediate couplings, all resources relax toward saturation as power laws, with saturation values and relaxation exponents decreasing monotonically with the coupling: an MBL-like behavior~\cite{Falcao2025NonstabilizernessDynamicsManyBody, FalcaoFaF2026} emerges without any quenched disorder.
At the largest couplings, the bootstrap classification favors unbounded, ultraslow double-logarithmic growth of the stabilizer and participation entropies, whereas the non-Gaussianity continues to relax toward saturation for both gauge groups.
We traced this law to the interplay of two gauge-invariance mechanisms: Gauss-law energetics, which forces a logarithmically growing perturbative depth in every superselection sector, and confinement, which restricts the accessible configuration space to a polynomially growing volume~\cite{SM}.
Gauge invariance thus emerges as an intrinsic mechanism constraining not only transport and entanglement but also the very generation of non-classicality.

Our analysis suggests that the Pauli weight remains confined to a polynomially growing set of strings organized by the configurational front~\cite{SM}, making this regime a natural target for truncated Pauli-propagation methods~\cite{Shao2025CharacterizingPauliPropagation, Martinez2025EfficientSimulationParametrized, Angrisani2025ClassicallyEstimatingObservables, Rudolph2025PauliPropagationComputational}, while the saturating NG indicates near-Gaussian dynamics amenable to Majorana-propagation techniques \cite{Miller2025SimulationFermionicCircuits}.
Finally, all three resources are measurable on current platforms~\cite{Oliviero2022MeasuringMagicQuantum, Haug2023ScalableMeasuresMagic}, making the ultraslow growth of complexity a concrete experimental target for quantum simulators of gauge theories, where these monotones quantify both the cost of classical simulation and the hardness of state preparation.

\bigskip
\footnotesize
\begin{acknowledgments}
\textit{Acknowledgments.}
G.C.~and J.C.H.~acknowledge funding by the Max Planck Society, the Deutsche Forschungsgemeinschaft (DFG, German Research Foundation) under Germany’s Excellence Strategy – EXC-2111 – 390814868, and the European Research Council (ERC) under the European Union’s Horizon Europe research and innovation program (Grant Agreement No.~101165667)—ERC Starting Grant QuSiGauge.
This work is part of the Quantum Computing for High-Energy Physics (QC4HEP) working group.
E.T. was funded by the Swiss National Science Foundation (SNSF) under Grant No. TMPFP2\_234754.
E.T. acknowledges CINECA (Consorzio Interuniversitario per il Calcolo Automatico) award, under the ISCRA initiative and Leonardo early access program, for the availability of high-performance computing resources and support.
\end{acknowledgments}
\normalsize
\bibliography{bibliography}

\begin{thebibliography}{128}
\bibitem [{Bas()}]{BasisNote}%
  \BibitemOpen
  \href@noop {} {\bibinfo {title} {Since the {SRE} and {PE} are basis-dependent, we evaluate them in the computational bases of the gauge-invariant formulations introduced below---the spin basis of the integrated {U}$(1)$ model and the dressed-site basis of the {SU}$(2)$ model---which are also the natural measurement bases in quantum simulations of these theories.}}\BibitemShut {Stop}%
\bibitem [{Sre()}]{SreNote}%
  \BibitemOpen
  \href@noop {} {\bibinfo {title} {We do not track the {SRE} for {SU}(2): in the dressed-site formulation the local dimension is $d\!=\!13$, and evaluating the {Pauli}-weight distribution over the $13^{{2L}}$ {Heisenberg}--{Weyl} strings is impractical already at ${L}\!=\!12$.}}\BibitemShut {Stop}%
\bibitem [{Ach()}]{AchieveNote}%
  \BibitemOpen
  \href@noop {} {\bibinfo {title} {Achievability is guaranteed by the binary matter occupations: hardcore particles cannot cross, so the $u$-th particle from the left of $\vb{z}_{0}$ must reach the $u$-th occupied site of $\vb{z}$, and some particle can always advance toward its target (in any chain of mutually blocking particles, the outermost one is free), so that no hop is ever wasted.}}\BibitemShut {Stop}%
\end{thebibliography}

\clearpage
\onecolumngrid
\section*{End Matter}
\twocolumngrid

\section*{Scaling of the saturating-power-law parameters with gauge coupling}
\label{sec:fit_parameter_scaling}
Another stage of the analysis of the saturating dynamics of quantum resources in \cref{fig_U1} and \cref{figSU2} concerns how the saturation value $X_\infty$ and the relaxation exponent $\alpha$ depend on the gauge coupling $g\!=\!\sqrt{g^2}$ of U$(1)$ and SU$(2)$, respectively.
Only traces classified as saturating in \cref{tab:time_fit_selection} enter this analysis: SRE$_2$ and PE$_2$ at $g^2\!\in\![4,12]$ and NG at $g^2\!\in\![4,20]$.
The weakest coupling $g^2=2$ is omitted in every panel: it saturates but lies in a distinct, fast-relaxing regime and is a clear outlier to the coupling trend (keeping it forces a spurious linear $X_\infty(g)$).
We compare five trend models:
\begin{equation}
  \begin{aligned}
    c & , & a & +bg, & a & +\frac{b}{g}, & a & g^{-b}, & a & +b e^{-g/g_0},
  \end{aligned}
  \label{eq_parameter_trends}
\end{equation}
ranked by \cref{eq_AIC} and weighted by the time-domain covariance errors; since those errors are inherited from the first stage, this analysis is exploratory.
We quantify the quality of each trend by the adjusted coefficient of determination
\begin{equation}
  R^{2}_{\mathrm{adj}}=1-(1-R^{2})\,\frac{n-1}{\,n-p-1\,},
  \label{eq_r2adj}
\end{equation}
where $R^{2}$ is given by \cref{eq_r2}, $n$ is the number of couplings entering the fit, and $p$ is the number of trend parameters; the adjustment penalizes $p$, so trends of different complexity in \cref{eq_parameter_trends} are compared fairly.

The selected trends and the corresponding $R^{2}_{\mathrm{adj}}$ are given in \cref{tab:u1_parameter_trends}, while the optimal fits are shown in \cref{figU1_parameters}.
The saturation value decreases monotonically with coupling: a power-law decay for SRE$_2$ and PE$_2$ and an exponential approach to a finite offset for NG, in all cases with $R^{2}_{\mathrm{adj}}\geq0.995$.
The relaxation exponent also decreases with coupling---linearly for SRE$_2$ and NG ($R^{2}_{\mathrm{adj}}=0.902$ and $0.988$)---signaling an ever-slower approach to the asymptote.
The PE$_2$ exponent is the least constrained ($R^{2}_{\mathrm{adj}}=0.737$ over five couplings) and its trend should not be over-interpreted.

\begin{table}[b]
  \caption{\textbf{U$(1)$ coupling-trend analysis.} Minimum-AIC trend among
    \cref{eq_parameter_trends} for the saturation value $X_\infty$ and the relaxation
    exponent $\alpha$, with the number of couplings $g$ entering each fit and the
    adjusted coefficient of determination $R^{2}_{\mathrm{adj}}$.}
  \label{tab:u1_parameter_trends}
  \begin{ruledtabular}
    \begin{tabular}{ccccc}
      Observable & Parameter  & \#pts & Selected trend        & $R^{2}_{\mathrm{adj}}$ \\
      \hline
      SRE$_2$    & $X_\infty$ & 5     & $31.8\,g^{-1.69}$     & 0.999                  \\
      PE$_2$     & $X_\infty$ & 5     & $17.3\,g^{-1.64}$     & 0.995                  \\
      NG         & $X_\infty$ & 7     & $4.0+41\,e^{-g/0.91}$ & 0.996                  \\
      SRE$_2$    & $\alpha$   & 5     & $0.44-0.10\,g$        & 0.902                  \\
      PE$_2$     & $\alpha$   & 5     & $0.70\,g^{-1.35}$     & 0.737                  \\
      NG         & $\alpha$   & 7     & $0.32-0.056\,g$       & 0.988                  \\
    \end{tabular}
  \end{ruledtabular}
\end{table}

\begin{figure*}[ht]
  \centering
  \includegraphics[width=\textwidth]{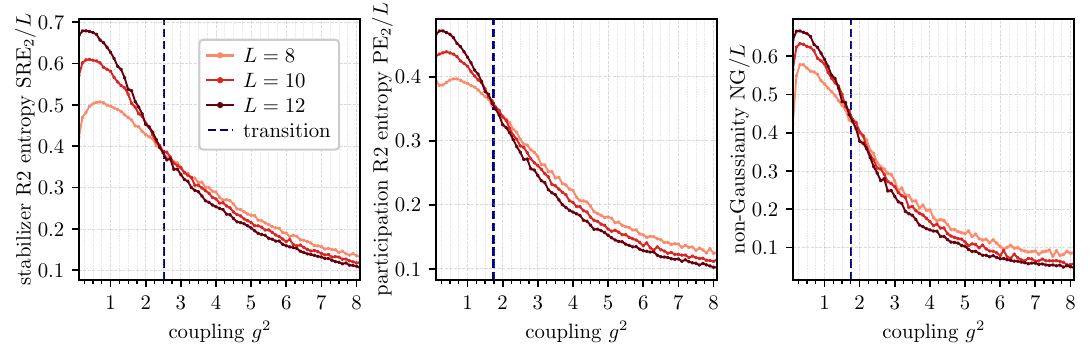}
\caption{\textbf{Finite-size scaling of static quantum resources in the U$(1)$ LGT.}
From left to right, the densities of stabilizer Rényi-2 entropy (SRE$_2/L$), participation Rényi-2 entropy (PE$_2/L$), and fermionic non-Gaussianity (NG$/L$) as functions of the gauge coupling $g^{2}$ for different lattice sizes $L=8,10,12$ (increasing color intensity).
The vertical dashed line marks the finite-size crossing point, separating the weak-coupling regime, where the densities grow with $L$, from the strong-coupling regime, where they decrease with $L$. In all cases, we averaged over $350$ background super-selection sectors.}
  \label{fig_U1_statics}
\end{figure*}

\section*{Finite-size scaling of static quantum resources}
The dynamical classification of \cref{tab:time_fit_selection} is intrinsically limited by the accessible time window: over any finite interval, a sufficiently slow unbounded growth is difficult to distinguish from an incomplete relaxation.
The static, equilibrium properties of the spectrum provide a complementary, time-independent diagnostic of the same physics.
Following the strategy used for the eigenstate entanglement entropy in the lattice Schwinger model~\cite{Jeyaretnam2025HilbertSpaceFragmentation}, we analyze the finite-size scaling of the \emph{densities} of the three quantum resources, $\mathrm{SRE}_2/L$, $\mathrm{PE}_2/L$, and $\mathrm{NG}/L$, shown in \cref{fig_U1_statics} as functions of the coupling $g^2$ for $L=8,10,12$.
The resources are evaluated on mid-spectrum eigenstates of the effective Hamiltonian of \cref{eq_U1_hamiltonian_integrated} and averaged over the background-charge superselection sectors entering \cref{eq_initial_state}, so that the static ensemble probes the same effective disorder that governs the quench dynamics of \cref{fig_U1}.

The two regimes are distinguished by the direction of the finite-size flow.
In the ergodic phase, mid-spectrum eigenstates are locally well described by Haar-random vectors in the gauge-constrained Hilbert space, for which all three resources are extensive with maximal densities~\cite{Turkeshi2025PauliSpectrumNonstabilizerness, Turkeshi2025MagicSpreadingRandom, Sierant2026FermionicMagicResources}: the densities then grow with $L$ toward their random-state values, as observed in \cref{fig_U1_statics} at weak coupling.
Conversely, in the localized regime eigenstates are related to computational-basis product states by quasilocal transformations~\cite{Abanin2019ColloquiumManybodyLocalization}, and Gauss-law fragmentation confines them to small Krylov fragments~\cite{Moudgalya2022QuantumManybodyScars, Jeyaretnam2025HilbertSpaceFragmentation}; the resources are then subextensive, and their densities \emph{decrease} with $L$, as observed at strong coupling.
The crossing point of the curves for successive system sizes therefore provides a finite-size estimate of the ergodicity-breaking coupling, marked by the dashed line in \cref{fig_U1_statics}, located at $g^2_c \approx 2$--$2.5$ and mutually consistent among the three observables.
 
Two consistency checks corroborate this estimate.
First, it matches the dynamical phenomenology of \cref{fig_U1}: for $g^2 \lesssim 2$ all resources equilibrate rapidly to near-maximal values, whereas the slow power-law and double-logarithmic regimes set in only for $g^2 \geq 4$, on the localized side of the crossing.
Second, the crossing is in quantitative agreement with the volume-to-area-law crossing of the mid-spectrum entanglement entropy in the same effective model, \cref{eq_U1_hamiltonian_integrated}, reported in Ref.~\cite{Jeyaretnam2025HilbertSpaceFragmentation} at a coupling that corresponds to $g^2 \approx 2.4$ in our convention.
Entanglement, nonstabilizerness, wavefunction delocalization, and non-Gaussianity thus locate the same ergodicity-breaking scale, reinforcing the conclusion of the main text that gauge invariance suppresses all complexity measures alike, rather than any single resource.
 
We emphasize the finite-size caveats, which parallel those discussed in Ref.~\cite{Jeyaretnam2025HilbertSpaceFragmentation}.
The crossing at $L \leq 12$ is a lower-bound estimate of the putative transition coupling and may drift with system size, as is well documented in disordered MBL~\cite{Sierant2025ManybodyLocalizationAge}; moreover, the available sizes cannot discriminate a genuine localization transition from a sharp crossover controlled by the residual, approximate fragmentation of the resonant graph at finite $g^{2}/w$ (cf.\ the SM  \cite{SM}).
Within these limitations, the static analysis independently confirms the existence of two regimes with opposite finite-size flows of the resource densities and places the dynamical slowdown of \cref{fig_U1} on the nonergodic side of the crossing.

\clearpage
\onecolumngrid
\begin{center}
  \textbf{\large Supplemental Online Material for \\"Quantum Resources in Disorder-Free Localization Dynamics of Gauge Theories" }\\[5pt]
  \vspace{0.1cm}
  \begin{quote}
{\small In this Supplemental Material, we detail the derivation of the effective models, the gauge-invariant construction of the fermionic non-Gaussianity, the statistical analysis of the resource dynamics, and the configuration-space argument for the ultraslow growth of quantum resources.}\\[10pt]
  \end{quote}
\end{center}
\setcounter{equation}{0}
\setcounter{figure}{0}
\setcounter{table}{0}
\setcounter{page}{1}
\setcounter{section}{1}
\makeatletter
\renewcommand{\theequation}{S\arabic{equation}}
\renewcommand{\thefigure}{S\arabic{figure}}
\renewcommand{\thesection}{S\Roman{section}}
\renewcommand{\thepage}{\arabic{page}}
\renewcommand{\thetable}{S\arabic{table}}
\vspace{0cm}
\twocolumngrid
\normalsize
\section{\texorpdfstring{$1\!+\!1$D U$(1)$}{(1+1)D U(1)} gauge-field integration}
\label{sec:sm_u1integration}
In one spatial dimension, exploiting U$(1)$ Gauss law in \cref{eq_U1_gausslaw} and fixing the left boundary electric field to $\hat{E}_{0,1}=\bar{E}_0$, we can sequentially integrate the gauge fields out \cite{Hamer1997SeriesExpansionsMassive, Brenes2018ManyBodyLocalizationDynamics} by expressing
\begin{equation}
  \hat{E}_{\site,\site{+}1}
  =\bar{E}_{0}+\sum_{j=1}^{\site}\qty[q_{j}+\hpsi^{\dagger}_{j}\hpsi_{j}\!+\!\frac{(-1)^{j}-1}{2}],
  \label{eq_U1_gaugeintegration}
\end{equation}
so that the electric energy term of \cref{eq_U1_hamiltonian} becomes a nonlocal functional of the matter charges.
Correspondingly, the remaining link operators $\Apara$ in the hopping term can be eliminated by gauge-transforming the matter fields as
\begin{equation}
  \hpsi_{\site}\to\prod_{j=1}^{\site{-}1}\Apara_{j,j+1}^{\dagger}\hpsi_{\site},
  \qquad
  \hpsi_{\site}^\dagger\to
  \hpsi_{\site}^\dagger\prod_{j=1}^{\site{-}1}\Apara_{j,j+1}.
  \label{eq_U1_dressed_fermions}
\end{equation}
These operators satisfy the same anti-commutation relations as the original matter fields and make the hopping term purely fermionic:
\begin{math}
  \hpsi_{\site}^{\dagger}\Apara_{\site,\site{+}1}\hpsi_{\site{+}1}\to
  \hpsi_{\site}^{\dagger}\hpsi_{\site{+}1}.
\end{math}
The gauge field is therefore fully encoded in the electric-energy term, whose dependence on matter and background charges is fixed by Gauss's law.
Finally, by applying the Jordan-Wigner transformation to the fermion fields, for which $\hpsi_{\site}^\dagger\hpsi_{\site}\!=\!\qty(\Sz_{\site}+1)/2$, one obtains an effective \emph{long-range} spin Hamiltonian whose interactions grow linearly with the lattice size \cite{Brenes2018ManyBodyLocalizationDynamics}:
\begin{equation}
  \begin{split}
    \ham_{\mathrm{U(1)}}^{\{q_{\site}\!\}}\!=\! & -\!w\sum_{\site=1}^{L-1}\qty(\Sp_{\site}\Sm_{\site{+}1}\!+\!\hc)
    \!+\!\frac{\mass}{2}\sum_{\site=1}^{L}(-1)^{\site}\Sz_{\site}                                                  \\
                                                & +\frac{g^{2}}{2}\sum_{\site=1}^{L-1}
    \qty[\sum_{j=1}^{\site}\qty(q_{j}+\frac{\Sz_{j}+(-1)^{j}}{2})]^2,
  \end{split}
  \label{eq_U1_hamiltonian_integrated2}
\end{equation}
up to an additive constant.
\cref{eq_U1_hamiltonian_integrated2} reflects that, after integrating out the gauge fields, the resulting Hamiltonian dynamics depends (through the coupling $g$) explicitly on the set of static charges $\{q_{\site}\}$ placed on the lattice and forming one of the $\mathcal{N}_{\mathrm{SS}}$ specific U$(1)$ superselection sectors (SS).
Consequently, states belonging to different superselection sectors evolve under different effective Hamiltonians $\ham^{\{q_{\site}\!\}}$.
\section{\texorpdfstring{SU$(2)$}{SU(2)} LGT: chromoelectric basis and dressed-site formulation}
\label{sec:sm_su2}
We collect here the explicit form of the SU$(2)$ link operators entering \cref{eq_hamiltonian_SU2} and the truncation adopted in our simulations; extensive presentations can be found in Refs.~\cite{Zohar2015FormulationLatticeGauge, Cataldi2024Simulating2+1DSU2, Cataldi2026DisorderFreeLocalizationFragmentation}.
In the chromoelectric basis $\ket{\spin,\mL,\mR}$ of a link, the left and right rotation generators $\hat{L}^{\nu}$ and $\hat{R}^{\nu}$ entering the SU$(2)$ Gauss law in \cref{eq_SU2_gausslaw} act within a spin shell $\spin$ as
\begin{subequations}
  \label{eq_LR_rishon_rotations}
  \begin{align}
    \mel{\spin^{\prime} \mL^{\prime} \mR^{\prime}}{\hat{L}^{\nu}}{\spin \mL \mR }  & =\delta_{\spin,\spin^{\prime}}S^{(\spin)\nu}_{\mL^{\prime},\mL} \delta_{\mR^{\prime},\mR}\,, \\
    \mel{\spin^{\prime} \mL^{\prime} \mR^{\prime} }{\hat{R}^{\nu}}{\spin \mL \mR } & =\delta_{\spin,\spin^{\prime}}\delta_{\mL^{\prime},\mL} S^{(\spin)\nu}_{\mR^{\prime},\mR}\,,
  \end{align}
\end{subequations}
where $S^{(\spin)\nu}$ are the spin-$\spin$ $\mathfrak{su}(2)$ matrices, so that the electric energy $\casimir\!=\!\sum_{\nu}(\hat{R}^{\nu})^2\!=\!\sum_{\nu}(\hat{L}^{\nu})^2$ is diagonal and equals the quadratic Casimir $\spin(\spin{+}1)$.
The parallel transporter $\NApara$ is instead off-diagonal in $\spin$, with matrix elements fixed by Clebsch-Gordan coefficients $C_{\spin_{1},m_{1};\,\spin_{2},m_{2}}^{\spin_{3},m_{3}}$ \cite{Zohar2015FormulationLatticeGauge}:
\begin{equation}
  \label{eq_parallel_transporter}
  \mel{\spin^{\prime}\mL^{\prime}\mR^{\prime}}{\hat{U}^{\alpha\beta}}{\spin\mL \mR}
  \!=\!\sqrt{\frac{2\spin\!+\!1}{2\spin ^{\mathrlap{\prime}}\!+\!1}}\:
  \overline{C^{\spin,\mL}_{\jonehalf,\alpha;\,\spin^{\prime},\mL^{\prime}}}
  C^{\spin^{\prime},\mR^{\prime}}_{\jonehalf,\beta;\,\spin,\mR}\!.
\end{equation}
In our simulations, gauge links and background charges are truncated to the two smallest irreps, $\spin,\spin_{q}\!\in\!\{0,1/2\}$ (hardcore-gluon approximation \cite{Cataldi2024Simulating2+1DSU2, Calajo2024DigitalQuantumSimulation}), a regime already sufficient to host quantum many-body scarring \cite{Calajo2025QuantumManybodyScarring}, Hilbert-space fragmentation, and disorder-free localization \cite{Cataldi2026DisorderFreeLocalizationFragmentation}.
Within this truncation, the SU$(2)$ Gauss law is solved locally by a \emph{dressed-site} construction, which fuses the matter fields with the adjacent link halves into gauge-invariant local multiplets: the result is a defermionized qudit model with a $13$-dimensional single-site basis, whose explicit matrix elements and global symmetry sectors are detailed in Ref.~\cite{Cataldi2026DisorderFreeLocalizationFragmentation}.
\section{Fermionic non-Gaussianity for Lattice Gauge Theories}
For a fermionic many-body state conserving the total particle number, a natural diagnostic of Gaussianity is provided by the covariance matrix
\begin{equation}
  C_{\site,\site^{\prime}}=\avg{\hpsi_{\site}^{\dagger}\hpsi_{\site^{\prime}}}.
\end{equation}
For a pure fermionic Gaussian state, \idest{} a Slater determinant, Wick's theorem implies that all correlation functions are fully determined by $C$, which satisfies $C^{2}=C$.
Therefore, its eigenvalues are restricted to $0$ and $1$.
For a generic interacting pure state, instead, the eigenvalues of $C$ are $\nu_{j}\in[0,1]$, and we define the non-Gaussianity (NG) as
\begin{equation}
  \mathrm{NG}=\sum_{j}\qty[-\nu_{j}\log(\nu_{j})-(1-\nu_{j})\log(1-\nu_{j})],
  \label{eq_NG}
\end{equation}
which vanishes for a Slater determinant and is strictly positive otherwise.
In LGTs, however, bare fermions are not gauge-invariant observables.
Therefore, for $\site\!\neq\!\site^{\prime}$, the bare correlator $\avg{\hpsi_{\site}^{\dagger}\hpsi_{\site^{\prime}}}$ must be replaced by a gauge-dressed object in which charge or color is parallel-transported from $\site^{\prime}$ to $\site$ through a Wilson line.
In a U$(1)$ LGT, this leads to the gauge-invariant covariance matrix
\begin{equation}
  C_{\site,\site^{\prime}}^{\mathrm{U(1)}}=
  \avg{
    \hpsi_{\site}^{\dagger}
    \qty[\prod_{j=\site}^{\site^{\prime}-1}\Apara_{j,j+1}]
    \hpsi_{\site^{\prime}}
  }.
  \label{eq_U1_FNG}
\end{equation}
In an SU$(2)$ LGT, the Wilson line carries color indices.
Formally, the natural gauge-covariant object is a $2L\!\times\!2L$ matrix, where, $\forall \alpha,\beta\in\qty{\rla,\gla}$:
\begin{equation}
  C_{(\site,\alpha),(\site^{\prime},\beta)}^{\mathrm{SU(2)}}=
  \avg{
    \hpsi_{\site,\alpha}^{\dagger}
    \qty[\mathcal{P}\prod_{j=\site}^{\site^{\prime}-1}\Apara_{j,j+1}]_{\alpha\beta}
    \hpsi_{\site^{\prime},\beta}
  }
\end{equation}
where $\mathcal{P}$ is a path ordering connecting the sites $\site,\site^{\prime}$ with the product of parallel transporters:
\begin{equation}
  \sum_{\substack{\gamma_{1},\dots,\\
      \gamma_{\site^{\prime}{-}\site{-}1}}}
  \Apara_{\site, \site{+}1}^{\alpha\gamma_{1}}
  \Apara_{\site{+}1, \site{+}2}^{\gamma_{1}\gamma_{2}}
  \cdots
  \Apara_{\site^{\prime}{-}2, \site^{\prime}{-}1}^{\gamma_{\site^{\prime}{-}\site{-}2}\gamma_{\site^{\prime}{-}\site{-}1}}
  \Apara_{\site^{\prime}{-}1, \site^{\prime}}^{\gamma_{\site^{\prime}{-}\site{-}1}\beta}.
\end{equation}
In both cases, diagonalizing the resulting covariance matrix yields eigenvalues $\nu_j\in[0,1]$, from which the fermionic non-Gaussianity is computed through \cref{eq_NG}.

In $1\!+\!1$D, gauge fields can be integrated out by Gauss' law (see \cite{Hamer1997SeriesExpansionsMassive, Brenes2018ManyBodyLocalizationDynamics} for U$(1)$ and \cite{Kuhn2015NonAbelianStringBreaking} for SU$(2)$), so that the covariance matrix becomes purely fermionic.
In the truncated SU$(2)$ \emph{dressed-site} formulation considered here \cite{Cataldi2024Simulating2+1DSU2, Calajo2024DigitalQuantumSimulation, Calajo2025QuantumManybodyScarring, Cataldi2026DisorderFreeLocalizationFragmentation}, one works instead in a local gauge-invariant basis in which Gauss' law is already satisfied site by site.
In this basis, the two endpoint color labels do not survive as independent physical fermionic modes.
As a consequence, the relevant covariance matrix is reduced to one effective fermionic mode per dressed site, \idest{} an $L\times L$ matrix (its practical implementation is available in \cite{Cataldi2026EdlgtExactDiagonalization}).

\section{Statistical analysis of quantum resources}
\label{sec:fit_selection}
We detail the statistical procedure used to classify the late-time growth of the resource time traces shown in \cref{fig_U1,figSU2}: the stabilizer Rényi-2 entropy (SRE$_2$), the participation Rényi-2 entropy (PE$_2$), and the fermionic non-Gaussianity (NG), for the U$(1)$ and SU$(2)$ theories.
For a fixed observable $X(t)$ and electric coupling $g^2$, the analysis asks whether the accessible time window resolves a relaxation to a finite value or remains compatible with continued ultraslow growth.
Namely, we compare the following four closed forms:
\begin{subequations}
  \label{eq_fit_models}
  \begin{align}
    X_{\mathrm{sat}}(t)    & = X_{\infty}-A\,t^{-\alpha}, \label{eq_fit_power}        \\
    X_{\log}(t)            & = a\log t + b, \label{eq_fit_log}                        \\
    X_{\log\log}(t)        & = a\log\log t + b, \label{eq_fit_loglog}                 \\
    X_{\mathrm{bridge}}(t) & = X_{\infty}-A\,(\log t)^{-\beta}. \label{eq_fit_bridge}
  \end{align}
\end{subequations}
Here $X_{\infty}$ is the asymptotic (saturation) value, $A\!>\!0$ an amplitude, $\alpha\!>\!0$ the relaxation exponent of the saturating power law, and $a,b$ the slope and intercept of the logarithmic laws.
\cref{eq_fit_power} describes a power-law relaxation toward the finite asymptote $X_{\infty}$, while \cref{eq_fit_log,eq_fit_loglog} describe unbounded logarithmic and double-logarithmic growth.
\cref{eq_fit_bridge} is a relaxation in logarithmic time with exponent $\beta\!>\!0$ that interpolates between the two limits: indeed, for small $\beta$,
\begin{equation}
  X_{\mathrm{bridge}}(t)=(X_{\infty}-A)+A\,\beta\,\log\log t+\mathcal{O}(\beta^{2}),
  \label{eq_bridge_expansion}
\end{equation}
it reduces to double-logarithmic growth as $\beta\!\to\!0$ and to a sharp saturation for large $\beta$; the fitted $\beta$ therefore measures how far a trace sits between bounded and unbounded growth.
The comparison tests whether the available time interval resolves saturation or remains compatible with continued ultraslow growth.

To account for a direct comparison, for each observable and coupling all models are evaluated on exactly the same time samples, summarized in \cref{tab:fit_windows}.
\begin{table}[t]
  \caption{\textbf{Time-domain fitting windows.} The same window applies to every
    candidate model and, at fixed coupling, to every listed observable. $\Nsamples$
    is the number of samples in the window.}
  \label{tab:fit_windows}
  \begin{ruledtabular}
    \begin{tabular}{cccccc}
      Theory & size   & Observables         & $g^2$       & $[t_{\min},t_{\max}]$ & $\Nsamples$ \\
      \hline
      U$(1)$   & $L=16$ & SRE$_2$, PE$_2$, NG & $2$--$20$   & $[10^{2},10^{15}]$    & 260         \\
      SU$(2)$  & $L=12$ & PE$_2$, NG          & $0.5$--$15$ & $[10,10^{4}]$         & 55          \\
    \end{tabular}
  \end{ruledtabular}
\end{table}
For U$(1)$ we use weighted least squares with the stored error bars as the relative standard deviations $\sigma_i$; for SU$(2)$, where error bars are unavailable, we use unweighted least squares ($\sigma_i=1$).
For the saturating power law, we impose $X_{\infty}\geq0$, $A\geq0$, and $10^{-6}\leq\alpha\leq5$.
We choose the initial values $X_{\infty}^{(0)}=1.05\max_t X(t)$ and $\alpha^{(0)}=0.2$.
We initialize the amplitude as follows.
\begin{equation}
  A^{(0)}=\max\!\left\{\left[X_{\infty}^{(0)}-X(t_{\min})\right]t_{\min}^{\alpha^{(0)}},10^{-8}\right\}.
\end{equation}
The logarithmic laws are initialized by linear regression in $\log t$ or $\log\log t$ and are otherwise unconstrained.
Parameter uncertainties are taken from the square roots of the diagonal of the least-squares covariance matrix; they quantify the local uncertainty within a fixed model and window and do not include the systematic uncertainty of model selection or window choice.
\subsection{Model-selection criterion}
For a fit with $\Nsamples$ samples the (weighted) residual sum of squares of model $m$ is
\begin{equation}
  \mathrm{RSS}_m=\sum_{i=1}^{\Nsamples}\left[\frac{X(t_i)-X_m(t_i)}{\sigma_i}\right]^2 .
  \label{eq_rss}
\end{equation}
Because the time grid is logarithmic and the disorder-averaged traces have correlated residuals, we do not rely on a single in-window criterion.
We classify each trace with three complementary diagnostics and report the quantities defined below.
\paragraph{(i) Local logarithmic slope.}
We define the local slope in logarithmic time, estimated by a local weighted quadratic regression in $\log t$, which has distinct slope decays for the different candidate laws:
\begin{equation}
  s(t)\equiv\frac{dX}{d\log t}=
  \begin{cases}
    a                    & \; \text{for log}         \\
    a/\log t             & \; \text{for loglog}      \\
    A\alpha\,t^{-\alpha} & \; \text{for saturation},
  \end{cases}
  \label{eq_slope_forms}
\end{equation}
so the product $s(t)\log t$ is asymptotically constant for double-logarithmic growth, decays to zero for saturation, and grows for a single logarithm.
This is completely independent of the asymptote $X_\infty$ and of additive offsets.
We fit $s(t)$ to the three forms in \cref{eq_slope_forms} and record the class with the largest coefficient of determination,
\begin{equation}
  \begin{aligned}
    R^{2}                      & \!=\!1\!-\!\frac{\mathrm{SS}_{\mathrm{res}}}{\mathrm{SS}_{\mathrm{tot}}}, &
    \mathrm{SS}_{\mathrm{res}} & \!=\!\!\sum_i\!\qty[y_i\!-\!\hat y_i]^2,                                  &
    \mathrm{SS}_{\mathrm{tot}} & \!=\!\!\sum_i\!\qty[y_i\!-\!\bar y]^2,
  \end{aligned}
  \label{eq_r2}
\end{equation}
where $y_i$ is the fitted quantity, $\hat y_i$ the model prediction, and $\bar y$ the mean of $\{y_i\}$; $\mathrm{SS}_{\mathrm{res}}$ and $\mathrm{SS}_{\mathrm{tot}}$ are the residual and total sums of squares.

\paragraph{(ii) Forward cross-validation.}
We split the window at a fraction $f$ of the samples into an early ``training'' segment and a late ``test'' segment, fit each model on the training segment, and evaluate the held-out error
\begin{equation}
  E^{(m)}_{\mathrm{CV}}=\sqrt{\frac{1}{N_{\mathrm{test}}}\sum_{i\in\text{test}}
    \left[\frac{X(t_i)-X_m(t_i)}{\sigma_i}\right]^2},
  \label{eq_cv}
\end{equation}
averaged over $f\in\{0.5,0.6,0.7,0.8\}$.
Saturation and continued growth coincide on the training segment and diverge on the held-out late points, so $E^{(m)}_{\mathrm{CV}}$ measures directly which continuation the data support.

\paragraph{(iii) Block bootstrap.}
To attach a confidence to the classification under correlated residuals, we generate $B$ surrogate traces by resampling the fit residuals in contiguous blocks and adding them to the fitted curve, repeat the cross-validation \eqref{eq_cv} on each, and define the class probability
\begin{equation}
  P_m=\frac{1}{B}\sum_{b=1}^{B}\mathbb{1}\!\left[\,m=\arg\min_{m'}E^{(m')}_{\mathrm{CV},\,b}\,\right],
  \label{eq_classprob}
\end{equation}
i.e.\ the fraction of resamplings in which model $m$ best predicts the held-out late times.
$P_m$ is the primary statistic reported in \cref{tab:time_fit_selection}.

\paragraph{(iv) Information criterion (cross-check).}
As an independent check, we also rank the models by the corrected Akaike information criterion (AIC),
\begin{equation}
  \mathrm{AIC}_m=\Nsamples\log\!\left(\frac{\mathrm{RSS}_m}{\Nsamples}\right)
  +2p+\frac{2p(p+1)}{\Nsamples-p-1},
  \label{eq_AIC}
\end{equation}
where $p$ counts the curve parameters plus the common residual scale estimated from the data, and the additive constant common to all models is omitted.
To compare each model with the best one, we introduce
\begin{equation}
  \begin{aligned}
    \Delta_m & \!=\!\mathrm{AIC}_m\!-\!\min_{m'}\mathrm{AIC}_{m'}          &
    W_m      & \!=\!\frac{e^{-\Delta_m/2}}{\sum_{m'}e^{-\Delta_{m'}\!/2}}.
  \end{aligned}
  \label{eq_DAIC}
\end{equation}
In detail, $\Delta_m\lesssim2$ indicates models that cannot be distinguished, $\Delta_m\gtrsim10$ indicates very little support.
Because neighboring samples are correlated, the nominal $\Nsamples$ overstates the information content; we therefore also evaluate \cref{eq_AIC} with an effective sample size
\begin{equation}
  \Nsamples^{\mathrm{eff}}=\Nsamples\,\frac{1-c_{1}}{1+c_{1}},
  \label{eq_neff}
\end{equation}
with $c_{1}$ the (window-dependent) lag-1 autocorrelation of the residuals, and use the criterion only as a consistency check on the diagnostics (i)--(iii).

\subsection{Model-selection results}
The outcome of the comparison is summarized in \cref{tab:time_fit_selection}.
The "selected" model $m$ is the one with the largest bootstrap probability $P_m$ defined in \eqref{eq_classprob}, and the verdict distinguishes a confident classification from an \emph{unresolved} one, in which the saturating and double-logarithmic fits differ by less than the statistical error over the entire window and the diagnostics no longer separate them.
\begin{table}
  \caption{\textbf{Time-domain model selection.} For each theory, observable, and coupling, we report the law preferred by the bootstrap \eqref{eq_classprob}, its probability $P$, and the verdict. ``sat.'' denotes the saturating power law
    \eqref{eq_fit_power} and ``$\log\log$'' the double-logarithmic law
    \eqref{eq_fit_loglog}.}
  \label{tab:time_fit_selection}
  \begin{ruledtabular}
    \begin{tabular}{ccccc}
      Theory & observable & $g^2$         & selected   & $P$            \\
      \hline
      U$(1)$   & SRE$_2$    & $4$--$12$     & sat.       & $1.00$         \\
      U$(1)$   & SRE$_2$    & $16,\,20$     & $\log\log$ & $0.95,\,0.94$  \\
      U$(1)$   & PE$_2$     & $4$--$12$     & sat.       & $1.00$         \\
      U$(1)$   & PE$_2$     & $16$--$20$    & $\log\log$ & $0.90,\,0.94$  \\
      U$(1)$   & NG         & $4$--$20$     & sat.       & $1.00$         \\
      \hline
      SU$(2)$  & PE$_2$     & $0.5,\,2.92$  & sat.       & $1.00$         \\
      SU$(2)$  & PE$_2$     & $5.33$        & $\log\log$ & $0.79$         \\
      SU$(2)$  & PE$_2$     & $7.75$        & $\log\log$ & $0.95$         \\
      SU$(2)$  & PE$_2$     & $10.17$--$15$ & $\log\log$ & $0.59$--$0.73$ \\
      SU$(2)$  & NG         & $0.5$--$15$   & sat.       & $1.00$         \\
    \end{tabular}
  \end{ruledtabular}
\end{table}
For U$(1)$, the three diagnostics agree on saturation for $g^2\!\leq\!12$: the slope $s(t)\log t$ decays, the saturating law minimizes $E_{\mathrm{CV}}$, $P_{\rm sat}\!=\!1$, and \cref{eq_AIC} excludes both logarithmic laws by large margins.
At $g^2\!\geq\!16$, the saturating and double-logarithmic fits become nearly indistinguishable within the window: the bootstrap favors the double-logarithmic law both for SRE$_2$ ($P\!=\!0.95,0.94$) and for PE$_2$ ($P\!=\!0.90,0.94$) at $g^2\!=\!16,20$.
Conversely, NG relaxes to saturation across the whole range ($P_{\rm sat}\!=\!1$).
The single logarithm \eqref{eq_fit_log} is never supported.
In \cref{fig_U1} the double-logarithmic alternative is overlaid (dashed red) on the curves where it remains compatible---SRE$_2$ and PE$_2$ at $g^2\!=\!16,20$---and the saturating power law (dashed black) is shown for all the other curves.

For SU$(2)$, NG saturates at every coupling ($P_{\rm sat}=1$).
PE$_2$ saturates at the weakest couplings ($g^2\leq2.92$) and is best described by double-logarithmic growth at stronger couplings ($g^2\geq5.33$, $P$ up to $0.95$).
The accessible window spans only $\sim3$ decades, which limits the discrimination, so the strong-coupling PE$_2$ classification, while favoring double-logarithmic growth, remains compatible with a continued slow approach to saturation.

These conclusions identify the best description within the simulated window and candidate set; they do not establish the $t\rightarrow\infty$ asymptotic law, which would require longer simulations and a stability analysis under shifts of both fitting endpoints.

\section{Configuration-space argument for the double-logarithmic growth of quantum resources}
In this section we derive controlled bounds on the time growth of the participation Rényi-2 entropy ($\PE$) and the stabilizer Rényi-2 entropy ($\SRE$) in the strong-coupling regime of the $(1{+}1)$D U$(1)$ lattice gauge theory (LGT), and we use them to explain the ultraslow, double-logarithmic growth reported in \cref{fig_U1} of the main text.
Both quantities are defined in the main text; the key observation exploited here is that each of them is the Rényi-2 entropy of a probability distribution generated by the dynamics --- for the $\PE$, the weight of the state over the computational basis, and for the $\SRE$, the weight of its expansion in Pauli strings.

The argument proceeds in three logically independent steps, which we outline here before developing them in turn; the precise definitions are introduced where each step is carried out.
\begin{itemize}
\item[(i)] \emph{Kinematics (exact).}
We first show, by linear algebra alone and for an \emph{arbitrary} state, that each entropy is bounded by the logarithm of how many basis configurations the state explores: the $\PE$ by the number of populated computational-basis states, and the $\SRE$ by the number of distinct \emph{differences} between such states.
No dynamics or coupling regime enters at this stage.
\item[(ii)] \emph{Dynamics (strong coupling).}
We then estimate how fast the accessible set grows in time.
For strong couplings $g^{2}$, every matter-hop costs an electric energy of $\Order{g^{2}}$, far above the hopping scale $w$.
Since the total energy is conserved up to $\Order{w}$, the state can permanently rearrange only into configurations with the same electric energy as the initial one, and can reach them only through high-order virtual processes.
Degenerate perturbation theory shows that each additional hop multiplies the required time by $g^{2}/w$, so the number of hops performed by time $t$---and with it the accessible neighborhood of the initial configuration---grows only \emph{logarithmically}.
\item[(iii)] \emph{Geometry (verified fragmentation input).}
The time dependence of the entropies follows from how the size of that neighborhood $\Vres(r)$ grows with its radius $r$.
The single physical input is that gauge confinement prevents the number of mobile charges from growing with the depth, so the neighborhood grows only \emph{polynomially} in the radius, $\Vres(r)\!\sim\! r^{k}$.
With an exact count of the resonant configurations via a transfer-matrix method up to $L\!=\!96$, the polynomial law is confirmed, the exponential alternative is excluded, and the exponent grows linearly with system size, $k(L)\simeq0.15L$, corresponding to a finite \emph{density} of mobile charges.
Combined with the logarithmic radius of step (ii), this yields the ultraslow double-logarithmic ceilings $\PE\lesssim k\log\log t$ and $\SRE\le2\PE\lesssim2k\log\log t$ --- the mechanism behind the slowdown observed in \cref{fig_U1,figSU2}.
\end{itemize}
\subsection{Rényi-2 entropies of dynamical distributions}
\label{sec:kinematics}
We write the time-evolved state of the $L$-site chain in the computational basis,
\begin{equation}
  \ket{\psi(t)}\!=\!\sum_{\vb{z}}C_{\vb{z}}(t)\,\ket{\vb{z}},
  \quad
  p_{\vb{z}}(t)\!=\!\abs{C_{\vb{z}}(t)}^{2},
  \quad
  \sum_{\vb{z}}p_{\vb{z}}(t)\!=\!1,
\end{equation}
where $\vb{z}=(z_{1},\dots,z_{L})$ with $z_{\site}\in\{0,\dots,d{-}1\}$ and total Hilbert-space dimension $D=d^{L}$.
Within a fixed superselection sector, the initial state of \cref{eq_initial_state} reduces to a single computational-basis product state, $\ket{\psi(0)}=\ket{\vb{z}_{0}}$, which we take as the starting point of the argument; the role of the sector average is discussed at the end of this section.
The two geometric objects controlling all bounds below are the \emph{support} and the \emph{difference set} of the computational distribution,
\begin{equation}
  \begin{aligned}
    \Om(t)  & =\{\vb{z}:p_{\vb{z}}(t)\ge p_{\mathrm{th}}\}, \\
    \dOm(t) & =\{\vb{z}\ominus\vb{z}':\vb{z},\vb{z}'\in\Om(t)\},
  \end{aligned}
  \label{eq:supports}
\end{equation}
where $\ominus$ denotes component-wise subtraction modulo $d$.
Here $p_{\mathrm{th}}$ is a fixed threshold separating the dynamically populated configurations from the perturbatively suppressed tail; the bounds below control the entropies of the weight inside $\Om(t)$, up to corrections that vanish with the total tail weight.

\paragraph{Participation entropy.}
The participation Rényi-2 entropy defined in the main text is, by construction, the Rényi-2 entropy of $p_{\vb{z}}$,
\begin{equation}
  \PE(t)=-\log\sum_{\vb{z}}p_{\vb{z}}(t)^{2}\equiv-\log\IPR(t),
  \label{eq:pe-def}
\end{equation}
where $\IPR\!=\!\sum_{\vb{z}}p_{\vb{z}}(t)^{2}$ is the inverse participation ratio~\cite{Liu2025QuantumAlgorithmsInverse}.
By the Cauchy--Schwarz (convexity) inequality
\begin{equation}
  \sum_{\vb{z}\in\Om}p_{\vb{z}}^{2}\ge\frac{\bigl(\sum_{\vb{z}\in\Om}p_{\vb{z}}\bigr)^{2}}{\abs{\Om}},
\end{equation}
so that
\begin{equation}
  \PE(t)\le\log\abs{\Om(t)}
  \label{eq:pe-bound}
\end{equation}

\paragraph{Pauli-weight distribution.}
For the qudit Pauli (Heisenberg--Weyl) group we write a string as $\Phat=\Z^{\vb u}\X^{\vb v}$, with $\vb u,\vb v\in\{0,\dots,d{-}1\}^{L}$ and $\omega=e^{2\pi i/d}$, whose action on the basis is
\begin{equation}
  \X^{\vb v}\ket{\vb{z}}=\ket{\vb{z}\oplus\vb v},
  \qquad
  \Z^{\vb u}\ket{\vb{z}}=\omega^{\vb u\cdot\vb{z}}\ket{\vb{z}},
\end{equation}
so that $\X^{\vb v}$ permutes basis states ($\oplus$ is addition mod $d$) while $\Z^{\vb u}$ is diagonal.
Following the standard Pauli/Bell-sampling normalization we define the \emph{Pauli-weight distribution}
\begin{equation}
  \Xi_{\Phat}(t)=\frac{1}{D}\,\abs{\mel{\psi(t)}{\Phat}{\psi(t)}}^{2},
  \label{eq:xi-def}
\end{equation}
which is a probability distribution on the Pauli group because purity gives $\sum_{\Phat}\Xi_{\Phat}\!=\!\tfrac1D\sum_{\Phat}\abs{\avg{\Phat}}^{2}\!=\!\Tr\rho^{2}\!=\!1$.
The stabilizer Rényi-2 entropy defined in the main text is then \emph{exactly} the Rényi-2 entropy of $\Xi$, shifted by $\log D$:
\begin{equation}
  \begin{aligned}
    \SRE(t) & \!=\!-\!\log\!\qty[\!\sum_{\Phat}\frac{\avg{\Phat(t)}^{4}}{D}]\!\!\!=\!\!\underbrace{-\log\sum_{\Phat}\Xi_{\Phat}(t)^{2}}_{\textstyle S_{2}^{\Xi}\,\ge\,0}\!-\!\log D.
    \label{eq:sre-renyi}
  \end{aligned}
\end{equation}
The offset $-\log D$ subtracts the stabilizer baseline: for a stabilizer state, $\Xi$ is uniform over its $D$-element stabilizer group, so $S_{2}^{\Xi}=\log D$ and $\SRE=0$; magic measures the spread of $\Xi$ \emph{beyond} that group.

\paragraph{Support of the Pauli weight.}
The relevant matrix element is
\begin{equation}
  \mel{\psi}{\Z^{\vb u}\X^{\vb v}}{\psi}
  =\sum_{\vb{z}}C^{*}_{\vb{z}\oplus\vb v}\,C_{\vb{z}}\,\omega^{\vb u\cdot(\vb{z}\oplus\vb v)},
  \label{eq:matrixelement}
\end{equation}
which is nonzero only if there exists $\vb{z}$ with $\vb{z},\,\vb{z}\oplus\vb v\in\Om(t)$, i.e.\ only if the $\X$-pattern is an admissible difference,
\begin{equation}
  \vb v\in\dOm(t).
  \label{eq:v-in-dOm}
\end{equation}
For each such $\vb v$ the $\Z$-pattern $\vb u$ ranges over at most $D$ values, so the number of Pauli labels carrying nonzero weight obeys $\abs{\supp\Xi}\le D\,\abs{\dOm(t)}$.
Inserting $S_{2}^{\Xi}\le S_{0}^{\Xi}=\log\abs{\supp\Xi}$ into \cref{eq:sre-renyi}, the $\log D$ cancels and we obtain the second exact bound,
\begin{equation}
  \SRE(t)\le\log\abs{\dOm(t)}. 
  \label{eq:sre-bound}
\end{equation}
Thus, the participation entropy is controlled by the support $\Om$, while the stabilizer Rényi entropy is controlled by its \emph{difference set} $\dOm$.

\paragraph{Exact marginal identity.}
Summing $\Xi$ over the $\Z$-pattern $\vb u$ and using the Plancherel relation $\sum_{\vb u}\omega^{\vb u\cdot(\vb{z}-\vb{z}')}\!=\!D\,\delta_{\vb{z},\vb{z}'}$ gives
\begin{equation}
  \sum_{\vb u}\Xi_{\Z^{\vb u}\X^{\vb v}}=\chi(\vb v)\equiv\sum_{\vb{z}}p_{\vb{z}}\,p_{\vb{z}\oplus\vb v},
  \label{eq:marginal}
\end{equation}
i.e.\ the $\vb v$-marginal of the Pauli weight is exactly the autocorrelation of the computational distribution.
Hence, the spread of $\Xi$ in $\vb v$ is \emph{locked} to the spread of $p_{\vb{z}}$ --- in particular $\chi(\vb 0)=\IPR=e^{-\PE}$ --- while all genuinely stabilizer-resolving content of the $\SRE$ lives in the $\vb u$-resolution \emph{within} each $\vb v$-block, exactly the part absorbed by $-\log D$ in \cref{eq:sre-renyi}.
This identity is the precise sense in which $\PE$ and $\SRE$ are driven by the \emph{same} distributional spread, and it underlies the common ultraslow functional form of the two resources discussed below.

\paragraph{Exact inequality $\SRE\le2\,\PE$.}
The marginal identity yields a theorem directly connecting the two resources.
Applying the Cauchy--Schwarz inequality within each $\vb v$-block ($\vb u$ runs over $D$ values) and then keeping only the $\vb v=\vb 0$ term, for which $\chi(\vb 0)=\IPR$,
\begin{equation}
  \sum_{\Phat}\Xi_{\Phat}^{2}
  =\sum_{\vb v}\sum_{\vb u}\Xi_{\Z^{\vb u}\X^{\vb v}}^{2}
  \ge\frac{1}{D}\sum_{\vb v}\chi(\vb v)^{2}
  \ge\frac{\IPR^{2}}{D}.
\end{equation}
Taking $-\log$ and subtracting $\log D$ as in \cref{eq:sre-renyi} gives
\begin{equation}
  \SRE(t)\;\le\;2\,\PE(t)
  \label{eq:sre-le-2pe}
\end{equation}
for \emph{any} pure state.
Every ceiling derived below for the participation entropy therefore automatically transfers to the stabilizer entropy, with at most twice the slope and no further assumptions.
\subsection{Strong-coupling dynamics and the logarithmic perturbative depth}
\label{sec:dynamics}
The bounds \cref{eq:pe-bound,eq:sre-bound} are kinematic: they hold for any state.
The dynamics enters only through the growth of $\Om(t)$ and $\dOm(t)$, which we now estimate in the strong-coupling regime.

After integrating out the gauge field in a fixed superselection sector $\{q_{\site}\}$, the dynamics is governed by the long-range spin Hamiltonian of the main text, which we split into a diagonal part and the hopping,
\begin{equation}
  \ham=\ham_{0}+\Vhat,
  \qquad
  \Vhat=-w\sum_{\site=1}^{L-1}\bigl(\Sp_{\site}\Sm_{\site{+}1}+\hc\bigr),
  \label{eq:H-split}
\end{equation}
where the diagonal part reads:
\begin{equation}
  \ham_{0}=\frac{m}{2}\sum_{\site=1}^{L}(-1)^{\site}\Sz_{\site}
  +\frac{g^{2}}{2}\sum_{\site=1}^{L-1}\eleE_{\site,\site{+}1}^{2},
  \label{eq:H0}
\end{equation}
with $\eleE_{\site,\site{+}1}$ being the electric field in \cref{eq_U1_gaugeintegration}.
We work in the strong-coupling regime $g^{2}\gg w$, where the relevant large parameter is $g^{2}/w$.

\paragraph{Gauss-law energetics of a hop.}
A hop on bond $(\site,\site{+}1)$ moves one fermion between the two \emph{sites} $\site$ and $\site{+}1$, changing the two site occupations as $\Sz_{\site}\!:-1\!\to\!+1$ and $\Sz_{\site{+}1}\!:+1\!\to\!-1$ (or vice versa): the total charge is conserved but displaced by one site.
Via Gauss's law, the gauge field $\eleE_{\site^{\prime},\site^{\prime}{+}1}=\bar E_{0}+\sum_{j\le\site^{\prime}}\bigl[q_{j}+(\Sz_{j}+(-1)^{j})/2\bigr]$ accumulates all the charges to the left of link $(\site^{\prime},\site^{\prime}{+}1)$, so a link with $\site^{\prime}<\site$ contains neither of the two sites and is untouched, while a link with $\site^{\prime}\ge\site{+}1$ contains both, and the two opposite occupation changes \emph{cancel} in the sum.
Only the traversed link $\site^{\prime}=\site$ contains site $\site$ but not site $\site{+}1$; hence, a hop changes a single link field:
\begin{equation}
\begin{aligned}
\eleE_{\site,\site{+}1}&\!\to\!\eleE_{\site,\site{+}1}\pm1,\\
  \eleE_{\site^{\prime},\site^{\prime}{+}1}&\!\to\!\eleE_{\site^{\prime},\site^{\prime}{+}1}\quad \forall \site^{\prime}\!\neq\!\site.
\end{aligned}
  \label{eq:single-link}
\end{equation}
The associated change of electric energy reads
\begin{equation}
  \Delta\mathcal{E}_{\pm}=\frac{g^{2}}{2}\bigl[(\eleE_{\site,\site{+}1}\pm1)^{2}-\eleE_{\site,\site{+}1}^{2}\bigr]
  =\frac{g^{2}}{2}(\pm2\eleE_{\site,\site{+}1}+1),
  \label{eq:hop-cost}
\end{equation}
and, since $\eleE_{\site,\site{+}1}\!\in\!\mathbb{Z}$, never vanishes: an elementary hop costs energy $\Order{g^{2}}$.
Consequently, the diagonal spectrum of $\ham_{0}$ splits into towers separated by gaps $\Order{g^{2}}$; since the total energy is conserved while the hopping $\Vhat$ can supply at most an energy of order $w\ll g^{2}$, weight can be transferred permanently only between configurations of the same tower---which we call \emph{resonant}---whereas inter-tower transitions remain virtual, suppressed by powers of $w/g^{2}$.

\paragraph{Perturbative depth.}
A configuration with $n$ hops away from $\ket{\vb{z}_{0}}$ can be reached only at order $n$ in degenerate perturbation theory: each of the $n$ successive hops contributes to a matrix element of order $w$, separated by $n{-}1$ off-resonant intermediate configurations, each contributing an energy denominator of order $1/g^{2}$ [\cref{eq:hop-cost}].
The effective amplitude of the process is therefore exponentially small in $n$,
\begin{equation}
  \ham^{[n]}_{\mathrm{eff}}=\Order{\,w\,(w/g^{2})^{\,n-1}\,}=\Order{\,w^{\,n}/g^{\,2(n-1)}\,}.
  \label{eq:Hn}
\end{equation}
An amplitude of this size transfers appreciable weight onto the target configuration only after a time of order its inverse, $t_{n}\sim1/\ham^{[n]}_{\mathrm{eff}}$, which grows exponentially with the order $n$.
We then define the \emph{perturbative depth} $\nstar(t)$ as the largest order resolved by time $t$, i.e., the largest $n$ such that $t_{n}\lesssim t$; inverting this relation yields
\begin{equation}
  t_{n}\sim\frac{1}{w}\Bigl(\frac{g^{2}}{w}\Bigr)^{\,n-1}
  \quad\Longrightarrow\quad
  \nstar(t)\sim\frac{\log(wt)}{\log(g^{2}/w)},
  \label{eq:depth}
\end{equation}
so the depth explored up to time $t$ grows only \emph{logarithmically}: extending the reach by one more hop multiplies the required time by $g^{2}/w\gg1$.
In this precise sense, energy conservation turns the strong coupling into a \emph{logarithmic clock} on configuration space.
By time $t$, the wavefunction therefore occupies the depth-$\nstar(t)$ ball about $\ket{\vb{z}_{0}}$ on the resonant connectivity graph (vertices $=$ configurations, edges $=$ resonant hops), with leakage out of the initial energy shell suppressed by $(w/g^{2})^{\nstar}$.
Denoting by $\Vres(r)$ the number of resonant configurations within $r$ hops from $\ket{\vb{z}_{0}}$, we obtain 
\begin{equation}
  \begin{split}
      \abs{\Om(t)}\!\lesssim\!\Vres\bigl(\nstar(t)\bigr),\\
      \abs{\dOm(t)}\!\le\!\abs{\Om(t)}^{2}\!\lesssim\!\Vres\bigl(\nstar(t)\bigr)^{2},
  \end{split}
  \label{eq:vol}
\end{equation}
where the square arises from simple counting: every difference $\vb v=\vb{z}\ominus\vb{z}'$ is generated by at least one of the $\abs{\Om}^{2}$ ordered pairs $(\vb{z},\vb{z}')\in\Om\times\Om$, so the number of \emph{distinct} differences cannot exceed the number of pairs.
Taking logarithms, this square is the origin of the factor of two in the SRE slope, consistent with the exact inequality \cref{eq:sre-le-2pe}.
\cref{eq:pe-bound,eq:sre-le-2pe,eq:depth,eq:vol} reduce the problem to a single geometric question: how does the resonant volume $\Vres(r)$ grow with the radius $r$?
If it grows \emph{polynomially}, $\Vres(r) \sim r^{k}$, then
\begin{equation}
  \begin{split}
    \PE(t) &\le\log\abs{\Om(t)} \lesssim \log \Vres(\nstar(t)) \sim k\log\nstar(t)                                      \\
           & \sim\!k\log\log(wt)\!-\!k\log\log(g^{2}/w)\!+\!\mathrm{const},
  \end{split}
  \label{eq:pe-loglog}
\end{equation}
which gives a \emph{double logarithm} whose curves are \emph{shifted down} as the coupling increases, through the offset $-k\log\log(g^{2}/w)$ --- exactly the trend observed in \cref{fig_U1} of the main text.
By \cref{eq:sre-le-2pe}, the stabilizer entropy inherits the same law with at most twice the slope,
\begin{equation}
\begin{split}
  \SRE(t)&\!\le\!2\PE(t)\\
  &\!\lesssim\!2k[\log\log(wt)\!-\!\log\log(g^{2}/w)]\!+\!\mathrm{const}.
\end{split}
  \label{eq:sre-loglog}
\end{equation}
If instead the volume grows \emph{exponentially}, $\Vres(r)\sim e^{cr}$, $\PE(t)\lesssim c\,\nstar(t)\sim\log(wt)$.
Distinguishing the two scenarios is the crux of the argument, and it is a question the time-domain data cannot settle: over the accessible window, $\log\log t$ varies by less than a factor of two.
It can, however, be settled \emph{exactly} in configuration space, as we show next.
\subsection{Exact counting of the resonant volume}
\label{sec:sm_counting}
We now evaluate the resonant volume $\Vres(r)$ exactly, in order to discriminate between the polynomial and exponential scaling.
The key observation is that, in a fixed superselection sector $\{q_{\site}\}$ with integer boundary field $\bar E_{0}$, every computational basis state $\vb{z}$ carries a completely determined, integer-valued electric-field configuration $\vb{E}(\vb{z})\!=\!\bigl(E_{1,2}(\vb{z}),\dots, E_{L-1, L}(\vb{z})\bigr)$, with $E_{\site,\site{+}1}(\vb{z})$ the field on link $(\site,\site{+}1)$ [\cref{eq_U1_gaugeintegration}].
The map $\vb{z}\!\mapsto\!\vb{E}(\vb{z})$ is injective: the matter charges are recovered from Gauss law in ~\cref{eq_U1_gausslaw}: $Q_{\site}\!=\!E_{\site,\site{+}1}\!-\!E_{\site-1,\site}\!-\!q_{\site}$.
Counting configurations is therefore equivalent to counting electric field profiles, and both conditions defining the resonant volume become integer arithmetic on these profiles.

Consider the distance first.
A hop changes a single link by one unit [\cref{eq:single-link}], so connecting $\vb{z}_{0}$ to $\vb{z}$ takes at least $\lVert\vb{E}(\vb{z})-\vb{E}(\vb{z}_{0})\rVert_{1}$ hops; this minimal number is always achievable \cite{AchieveNote} and defines the distance
\begin{equation}
  \dist(\vb{z}_{0},\vb{z})=\sum_{\site=1}^{L-1}\abs{E_{\site,\site{+}1}(\vb{z})-E_{\site,\site{+}1}(\vb{z}_{0})}.
  \label{eq:distance}
\end{equation}
Consider next the energy.
The electric-energy offset between $\vb{z}$ and $\vb{z}_{0}$ is $\frac{g^{2}}{2}\Delta\varepsilon(\vb{z})$, with the integer energy mismatch $\Delta\varepsilon(\vb{z})=\sum_{\site}[E^{2}_{\site,\site{+}1}(\vb{z})-E^{2}_{\site,\site{+}1}(\vb{z}_{0})]$.
Since the total energy is conserved and the hopping can supply at most an energy of order $w$, the state can permanently occupy only configurations whose electric energy matches the initial one to within $w$; any nonzero mismatch costs at least $g^{2}/2\gg w$, so only the shell of \emph{resonant} (degenerate) configurations, \idest{} $\Delta\varepsilon(\vb{z})=0$, are accessible.
Moreover, each hop changes $\dist$ by exactly one and $\Delta\varepsilon$ by an odd integer [\cref{eq:hop-cost}]: both parities flip at every hop, both counts starting from zero, so they remain locked at all times, $\Delta\varepsilon\!\equiv\!\dist (\mathrm{mod}2)$; in particular, the resonant shell contains only even distances.
We define
\begin{equation}
  \Vres(r)=\#\qty{\vb{z}:\ \Delta\varepsilon(\vb{z})=0,\ \dist(\vb{z}_{0},\vb{z})\le r},
  \label{eq:vres-def}
\end{equation}
as the resonant volume, which counts the $\#$ of degenerate configurations ignoring their connectivity.
$\Vres$ is thus a rigorous upper bound on the dynamically accessible set, and an upper bound is all that the one-sided chain \cref{eq:pe-bound,eq:vol} requires.

\begin{figure}[t]
  \centering
  \includegraphics[width=1\columnwidth]{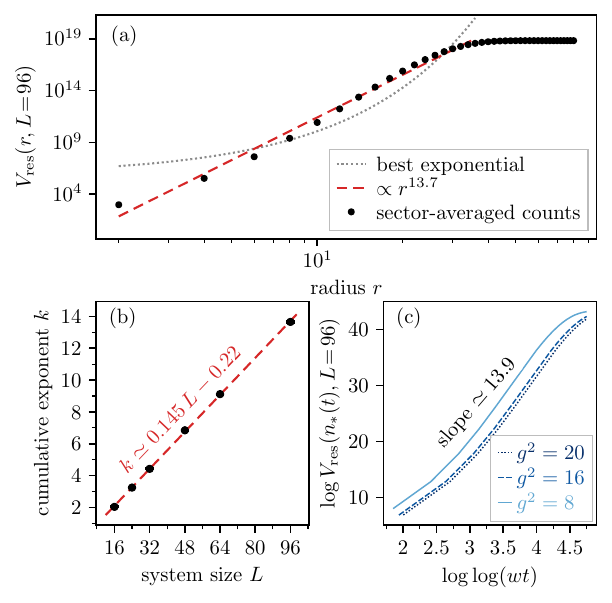}
\caption{\textbf{Exact counting of the resonant volume.}
(a) Sector-averaged resonant volume $\Vres(r)$ versus radius $r$ at $L\!=\!96$ (geometric mean of the exact counts over $500$ superselection sectors, dots): the average follows the power law $\Vres\!\sim\!r^{k}$ (dashed red), whereas the best exponential fit on the same window (dotted) bends away; the flattening at the largest $r$ marks the exhaustion of the finite shell.
(b) Exponent $k$ of the sector-averaged volume versus system size $L$ ($500$ sectors per size, bootstrap error bars); the dashed red line is the linear fit of \cref{eq:kL}.
(c) $\log\Vres(\nstar(t))$ versus $\log\log(wt)$, from the sector-averaged volume of (a), for increasing coupling $g^{2}$ (light to dark blue).}
  \label{fig_vres}
\end{figure}
\paragraph{Transfer-matrix evaluation.}
A brute-force evaluation of all the $\vb{z}$ configurations satisfying \cref{eq:vres-def} would enumerate every possible matter configuration, $\binom{L}{L/2}$, which already for $L\!=\!96$ counts $\sim\!10^{27}$ states---hopeless.
The count, however, can be organized as a single left-to-right scan of the chain, because all the ingredients of \cref{eq:vres-def} $(\vb{E}(\vb{z}),\Delta\varepsilon(\vb{z}), \dist(\vb{z}_{0},\vb{z}))$ are built up link by link: the field profile $\vb{E}(\vb{z})$ follows from Gauss's law, $E_{\site,\site{+}1}\!=\!E_{\site-1,\site}\!+\!Q_{\site}\!+\!q_{\site}$, while the distance $\dist(\vb{z}_{0},\vb{z})$ and the energy mismatch $\Delta\varepsilon(\vb{z})$ are sums of single-link contributions.

For a given fixed matter occupation of the first $s$ sites, the remaining sites can be completed based on only three integers: the current field $E_{s,s+1}$, the running mismatch $\sum_{\site\le s}[E_{\site,\site{+}1}^{2}\!-\!E^{2}_{\site,\site{+}1}(\vb{z}_{0})]$, and the running distance $\sum_{\site\le s}\abs{E_{\site,\site{+}1}\!-\!E_{\site,\site{+}1}(\vb{z}_{0})}$.
All the partial configurations sharing this same triplet admit exactly the same completions and can be merged into a single "state" that records only their multiplicity.
From site $s$ to site $s{+}1$, the vector of these multiplicities gets updated with the \emph{transfer-matrix} step: the two matter occupation choices on site $s{+}1$ map each triplet onto its updated triplets. 
After the last site, $\Vres(r)$ is read off by summing the multiplicities of all triplets with $\Delta\varepsilon\!=\!0$ and $\dist\!\leq\!r$.

The cost of the scan is controlled by the distance cap $r_{\max}$, an algorithmic cutoff independent of $L$: only triplets with running distance $\le r_{\max}$ are propagated, which is lossless for every radius $r\le r_{\max}$ in \cref{eq:vres-def}, since the running distance can only grow along the scan.
The cap also bounds the other two registers: every unit of field deviation on a link contributes one unit to the distance, so the fields range over $\Order{r_{\max}}$ values and the running mismatches over $\Order{r_{\max}^{2}}$.
The ledger therefore contains at most $\Order{r_{\max}^{4}}$ triplets, independently of $L$, and the full scan costs $\Order{L\,r_{\max}^{4}}$ operations---linear in system size at fixed cap.
For our simulations, we fix $r_{\max}=80$, far above the perturbative depths $\nstar(t)\lesssim10$ in \cref{eq:depth} reached at strong coupling within the simulated time window, and wide enough for $\Vres$ to grow by several orders of magnitude across the fitted range of distances $r$.
As for the multiplicities of the triplets, they are stored as arbitrary-precision integers (they exceed $10^{13}$), so the count is exact; the implementation was validated integer-for-integer against brute-force enumeration for $L\le16$.

Notice that, the transfer matrix method can be used within each fixed background-charge superselection sector contributing to the initial state \cref{eq_initial_state} of the main text.
The resulting $\Vres(r)$ prediction in the superposition of sector is obtained via sector-averaging.
In \cref{fig_vres}(a), we show the sector-averaged (over 500 drawn sectors) $\Vres$ for $L\!=\!96$, which clearly grows as a power law over nineteen orders of magnitude across the fitted window
\begin{equation}
  \Vres(r)\sim r^{k},\qquad k\simeq13.7\ \text{at}\ L=96.
  \label{eq:vres-powerlaw}
\end{equation}
An exponential growth bends visibly away from the exact counts in \cref{fig_vres}(a) and is excluded by the information criterion in \cref{eq_DAIC}, which favors the power law by $\Delta\mathrm{AIC}\!\simeq\!-2.2\times10^{4}$ (pooled over the $500$ sectors at $L\!=\!96$).
By repeating this protocol for $L\!=\!16$--$96$, we obtain what is shown in \cref{fig_vres}(b).
Namely, for each size $L$ we draw $500$ superselection sectors and, as in \cref{fig_vres}(a), extract the exponent $k$ from a linear fit of the logarithm of their geometric-mean volume versus $\log r$.
The exponent is \emph{extensive},
\begin{equation}
  k(L)\simeq 0.145\,L,
  \label{eq:kL}
\end{equation}
i.e., confinement bounds the \emph{density} of mobile charges, about $0.145$ per site, rather than their total number.
At $L\!=\!16$, the size simulated for the U(1) LGT in \cref{fig_U1}, this gives $k\approx2.0$ (the precise value depends mildly on the fit window), consistent with the $\Order{1}$ double-logarithmic slopes observed there.

At fixed $L$, $k$ enters the ceiling \cref{eq:pe-loglog} simply as a finite number.
As an example, in \cref{fig_vres}(c), we compute and show the scaling of $\log\Vres(\nstar(t))$ versus $\log\log(wt)$ for $w\!=\!1/2$ and different couplings $g^{2}$, evaluated on the sector-averaged volume of \cref{fig_vres}(a).
As expected from \cref{eq:pe-loglog}, we obtain approximately straight lines, with slope set by the exponent $k$ of \cref{eq:vres-powerlaw}, that shift rigidly downward by the offset $k\log\log(g^{2}/w)$ as the coupling grows.
This reproduces the trend of \cref{fig_U1}; the bending at the largest times occurs where $\Vres$ exhausts the finite $L=96$ shell.

Conversely, in the thermodynamic limit, the ceiling constrains resource \emph{densities}, $\PE/L\!\lesssim\!0.145\log\log(wt)\!+\!\mathrm{const}$, and the resonant volume $\Vres$ becomes a vanishingly small fraction of the total number of configurations: $\Vres/\binom{L}{L/2}\!\approx\!10^{-9}$ at $L\!=\!96$.
\end{document}